\documentclass[onecolumn]{IEEEtran}
\usepackage{graphicx} 

\usepackage[latin1]{inputenc}
\usepackage[english]{babel}
\usepackage[margin=1in]{geometry}

\usepackage{blindtext}
\usepackage{hyperref}
\hypersetup{
    colorlinks=true,
    linkcolor=blue,
    filecolor=magenta,      
    urlcolor=cyan,
}
 
\usepackage{amsthm}
\usepackage[normalem]{ulem}
\usepackage{amsmath}
\usepackage{amssymb}
\usepackage{mathrsfs}
\usepackage{booktabs}
\usepackage{color}
\usepackage{xcolor}
\usepackage{array}
\usepackage{cite}
\usepackage{soul}
\usepackage{mathtools}
\usepackage{bm}
\usepackage{multirow}
\usepackage{makecell}
\usepackage{thmtools}
\usepackage{thm-restate}
\usepackage{tikz}
\usepackage{adjustbox}
\usetikzlibrary{graphs,positioning}
\newtheorem{theorem}{Theorem}[section]

\newtheorem{lemma}[theorem]{Lemma}
\newtheorem{proposition}[theorem]{Proposition}
\newtheorem{construction}{Construction}

\newtheorem{example}{Example}
\newtheorem{remark}[theorem]{Remark}

\usepackage{cleveref}

\newcommand\nc\newcommand
\nc{\cA}{\mathcal{A}}\nc{\cB}{\mathcal{B}}\nc{\cC}{\mathcal{C}}\nc{\cD}{\mathcal{D}}
\nc{\cE}{\mathcal{E}}\nc{\cF}{\mathcal{F}}\nc{\cG}{\mathcal{G}}\nc{\cH}{\mathcal{H}}
\nc{\cI}{\mathcal{I}}\nc{\cJ}{\mathcal{J}}\nc{\cK}{\mathcal{K}}\nc{\cL}{\mathcal{L}}
\nc{\cM}{\mathcal{M}}\nc{\cN}{\mathcal{N}}\nc{\cO}{\mathcal{O}}\nc{\cP}{\mathcal{P}}
\nc{\cQ}{\mathcal{Q}}\nc{\cR}{\mathcal{R}}\nc{\cS}{\mathcal{S}}\nc{\cT}{\mathcal{T}}
\nc{\cU}{\mathcal{U}}\nc{\cV}{\mathcal{V}}\nc{\cW}{\mathcal{W}}\nc{\cX}{\mathcal{X}}
\nc{\cY}{\mathcal{Y}}\nc{\cZ}{\mathcal{Z}}

\nc{\bba}{\mathbf{a}}\nc{\bbb}{\mathbf{b}}\nc{\bbc}{\mathbf{c}}\nc{\bbd}{\mathbf{d}}
\nc{\bbe}{\mathbf{e}}\nc{\bbf}{\mathbf{f}}\nc{\bbg}{\mathbf{g}}\nc{\bbh}{\mathbf{h}}
\nc{\bbi}{\mathbf{i}}\nc{\bbj}{\mathbf{j}}\nc{\bbk}{\mathbf{k}}\nc{\bbl}{\mathbf{l}}
\nc{\bbm}{\mathbf{m}}\nc{\bbn}{\mathbf{n}}\nc{\bbo}{\mathbf{o}}\nc{\bbp}{\mathbf{p}}
\nc{\bbq}{\mathbf{q}}\nc{\bbr}{\mathbf{r}}\nc{\bbs}{\mathbf{s}}\nc{\bbt}{\mathbf{t}}
\nc{\bbu}{\mathbf{u}}\nc{\bbv}{\mathbf{v}}\nc{\bbw}{\mathbf{w}}\nc{\bbx}{\mathbf{x}}
\nc{\bby}{\mathbf{y}}\nc{\bbz}{\mathbf{z}}

\nc{\bbA}{\mathbf{A}}\nc{\bbB}{\mathbf{B}}\nc{\bbC}{\mathbf{C}}\nc{\bbD}{\mathbf{D}}
\nc{\bbE}{\mathbf{E}}\nc{\bbF}{\mathbf{F}}\nc{\bbG}{\mathbf{G}}\nc{\bbH}{\mathbf{H}}
\nc{\bbI}{\mathbf{I}}\nc{\bbJ}{\mathbf{J}}\nc{\bbK}{\mathbf{K}}\nc{\bbL}{\mathbf{L}}
\nc{\bbM}{\mathbf{M}}\nc{\bbN}{\mathbf{N}}\nc{\bbO}{\mathbf{O}}\nc{\bbP}{\mathbf{P}}
\nc{\bbQ}{\mathbf{Q}}\nc{\bbR}{\mathbf{R}}\nc{\bbS}{\mathbf{S}}\nc{\bbT}{\mathbf{T}}
\nc{\bbU}{\mathbf{U}}\nc{\bbV}{\mathbf{V}}\nc{\bbW}{\mathbf{W}}\nc{\bfX}{\mathbf{X}}
\nc{\bbY}{\mathbf{Y}}\nc{\bbZ}{\mathbf{Z}}

\nc{\sA}{\mathsf{A}}\nc{\sB}{\mathsf{B}}\nc{\sC}{\mathsf{C}}\nc{\sD}{\mathsf{D}}
\nc{\sE}{\mathsf{E}}\nc{\sF}{\mathsf{F}}\nc{\sG}{\mathsf{G}}\nc{\sH}{\mathsf{H}}
\nc{\sI}{\mathsf{I}}\nc{\sJ}{\mathsf{J}}\nc{\sK}{\mathsf{K}}\nc{\sL}{\mathsf{L}}
\nc{\sM}{\mathsf{M}}\nc{\sN}{\mathsf{N}}\nc{\sO}{\mathsf{O}}\nc{\sP}{\mathsf{P}}
\nc{\sQ}{\mathsf{Q}}\nc{\sR}{\mathsf{R}}\nc{\sS}{\mathsf{S}}\nc{\sT}{\mathsf{T}}
\nc{\sU}{\mathsf{U}}\nc{\sV}{\mathsf{V}}\nc{\sW}{\mathsf{W}}\nc{\sX}{\mathsf{X}}
\nc{\sY}{\mathsf{Y}}\nc{\sZ}{\mathsf{Z}}

\nc{\set}[1]{\llbracket #1 \rrbracket}

\newcommand{\bal}[1]{\begin{align}\label{#1}}
\newcommand{\eal}{\end{align}}

\renewcommand{\leq}{\leqslant}

\renewcommand{\geq}{\geqslant}

\renewcommand{\Bbb}{\mathbb}

\renewcommand{\Bbb}{\mathbb}

\newcommand{\Ftwo}{{{\Bbb F}}_{\!2}}

\theoremstyle{definition}
\newtheorem{definition}{Definition}[section]

\title{New Capacity Bounds for PIR on Graph and Multigraph-Based Replicated Storage}

\author{Xiangliang Kong,
    Shreya Meel,
    Thomas Jacob Maranzatto, 
    Itzhak Tamo, 
    and Sennur Ulukus 
\thanks{X. Kong (rongxlkong@gmail.com) and I. Tamo (tamo@tauex.tau.ac.il) are with the Department of Electrical Engineering-Systems, Tel Aviv University, Tel Aviv-Yafo 6997801, Israel.}
\thanks{S. Meel (smeel@umd.edu), T. J. Maranzatto (tmaran@umd.edu), and S. Ulukus (ulukus@umd.edu) are with the Department of Electrical and Computer Engineering, University of Maryland, College Park, MD, USA}
\thanks{The work of X. Kong and I. Tamo was supported  by the European Research Council (ERC) under Grant 852953.}
}

\begin{document}

\maketitle

\begin{abstract}
In this paper, we study the problem of private information retrieval (PIR) in both graph-based and multigraph-based replication systems, where each file is stored on exactly two servers, and any pair of servers shares at most $r$ files. We derive upper bounds on the PIR capacity for such systems and construct PIR schemes that approach these bounds.

For graph-based systems, we determine the exact PIR capacity for path graphs and improve upon existing results for complete bipartite graphs and complete graphs.

For multigraph-based systems, we propose a PIR scheme that leverages the symmetry of the underlying graph-based construction, yielding a capacity lower bound for such multigraphs. Furthermore, we establish several general upper and lower bounds on the PIR capacity of multigraphs, which are tight in certain cases.

\end{abstract}

\section{Introduction}

Introduced in~\cite{chor}, private information retrieval (PIR) is the problem of retrieving a specific file from a database stored across multiple servers without revealing the identity of the desired file to any of them, in the information-theoretic sense. The goal of PIR is to design retrieval schemes that minimize the total communication cost, i.e., the total number of bits exchanged between the user and the servers. This was the primary performance metric considered in~\cite{chor} and has remained a central focus in subsequent works; see, for example,~\cite{Ambainis97, BI01, BIKR02, Yekhanin08, DG16}.

Given the typical sizes of files users seek to access today, it is reasonable to assume that the size of the retrieved file is significantly larger than the number of bits uploaded by the user, i.e., the queries sent to the servers. This renders the upload cost negligible in comparison to the download cost from the servers~\cite{CHY15}. Motivated by this observation, Sun and Jafar studied the PIR problem in~\cite{SJ17}, where the objective is to minimize the download cost, i.e., the total number of bits sent by the servers.

In their framework, $K$ files are replicated across $N$ non-communicating (non-colluding) servers. The user generates and sends $N$ queries, one to each server. Upon receiving its query, each server truthfully responds with an answer, enabling the user to reconstruct the desired file. The \emph{rate} of a PIR scheme is defined as the number of desired message bits retrieved per downloaded bit, and the supremum of all achievable rates is referred to as the PIR \emph{capacity}.

Building on the characterization of PIR capacity in~\cite{SJ17}, a wide range of PIR variants have since been explored, motivated by diverse system assumptions and practical considerations. These studies have extended the classical PIR framework to more general and realistic settings. Notable examples include:
\begin{itemize}
    \item PIR with colluding servers, where subsets of servers may share their received queries in an attempt to infer the user's desired file~\cite{colluding, coded_colluding_2017, mdstpir}.
    \item Coded PIR, where files are not merely replicated but encoded using linear codes and distributed across servers~\cite{CHY15, BU18, TGE18, KLRA19, ZG19}.
    \item PIR under adversarial models, accounting for threats such as eavesdropping or Byzantine behavior by some servers~\cite{banawan_eaves, nan_eaves, byzantine_tpir}.
    \item Symmetric PIR (SPIR), which adds the constraint of database privacy, ensuring the user learns nothing beyond the requested file~\cite{c_spir,wang_spir, tspir_mdscoded}.
    \item Weakly private PIR, which allows for limited information leakage and investigates the trade-off between privacy and download efficiency~\cite{STL19, GZT20, LKRAY21_JSAIT, LKRAY21_TIT, QZTL22}.
    \item PIR with side information, including scenarios where the user knows parts of the database in advance~\cite{tpir_sideinfo, kadhe_singleserver_pir}, as well as cache-aided PIR, which leverages local caching to improve retrieval efficiency~\cite{wei_banawan_cache_pir}.
    \item PIR schemes that explore the trade-off between storage overhead and download cost~\cite{TSC18, Tian20, GZT21}.
\end{itemize}
These and many other extensions continue to be active areas of research. For a comprehensive overview of PIR and its numerous variants, we refer the reader to the survey~\cite{ulukusPIRLC}.

All of the aforementioned PIR models are considered over either fully replicated storage systems, where each server stores all files, or coded storage systems, where the data is distributed across servers using an error-correcting code. 

Although coding techniques for storage systems have seen significant theoretical and practical advances in recent years, many system designers continue to prefer replication over coding due to several advantages, including improved resiliency to data loss, implementation simplicity, ease of file updates, and high availability (see, e.g.,~\cite{cassandra20192,Apache_replication,ER10,SE15,YY19}). Moreover, given the vast volumes of data stored in modern applications, fully replicating an entire database across all servers is often impractical. Instead, in most large-scale storage systems, each server stores only a subset of the database (see, e.g.,~\cite{AKT20,WCJ19_ISIT,WCJ19_GLOBECOM}), and each file is replicated across a small number of servers.

This practical consideration motivated the study of the PIR problem in storage systems with limited replication using a graph-based model in~\cite{graphbased_pir}. In this model, a graph with $N$ vertices and $K$ (hyper-)edges represents a storage system, where each vertex corresponds to a server, and each (hyper-)edge represents a file and consists of the set of vertices (i.e., servers) that store it. This model includes the fully replicated storage system as a special case---namely, a replicated storage system represented by a hypergraph with $K$ multiple hyperedges, each of which consists of all the vertices. 

As in~\cite{SJ17}, the PIR problem over replicated storage systems modeled by graphs---referred to as \emph{the PIR problem for graphs}---was defined by Raviv, Tamo and Yaakobi in~\cite{graphbased_pir}, where they considered the PIR problem for simple graphs\footnote{In a simple graph, every edge consists of exactly two vertices, and every pair of vertices defines at most one edge.} under server collusion. Subsequent work in~\cite{BU19} further explored this problem by deriving bounds on the PIR capacity of certain regular graphs, where the PIR capacity of a graph is defined as the PIR capacity of the replicated storage system modeled by that graph. Later in~\cite{asymp_gxstpir}, the authors studied the PIR capacity of graphs in the asymptotic regime, where the number of files tends to infinity, while incorporating additional constraints such as server collusion and secure storage.

More recently,~\cite{SGT23} established new upper bounds on PIR capacities for several classes of graphs, including complete, star, and bipartite graphs. However, the exact PIR capacity for many of these graph families remains unknown, except for cases with a small number of vertices. For example, the PIR capacity of the star graph is still unknown beyond the case of $ K = 4 $, which was only recently resolved in~\cite{YJ23}.

In the same spirit as~\cite{graphbased_pir, BU19, SGT23, YJ23}, we further investigate the PIR problem for several important classes of graphs. Moreover, to overcome some of the limitations inherent in the case of simple graphs, we extend the study to multigraph-based storage systems with finite and uniform edge multiplicity $ r \geq 2 $. Specifically, our contributions are summarized as follows:
\begin{itemize}
    \item For \textbf{graph-based} systems, we consider the PIR problem for three classes of graphs: path graphs, complete bipartite graphs, and complete graphs. For path graphs, we show that the PIR capacity of a path graph with $N$ vertices is $2/N$. For complete bipartite graphs, we establish improved upper and lower bounds on the PIR capacity. Finally, for complete graphs, we present a PIR scheme that achieves the capacity for $N = 3$, and attains a higher rate than existing schemes for $N \geq 4$.
    
    \item For \textbf{multigraph-based} systems, we first propose a general PIR scheme construction based on schemes for graph-based systems that satisfies certain symmetry conditions. This construction yields a lower bound on the PIR capacity for certain classes of multigraphs. We also establish several general lower and upper bounds on the PIR capacity of multigraphs, and demonstrate that these bounds are tight for specific classes of multigraphs under certain parameter regimes.
\end{itemize}

For the reader's convenience, we summarize our results on the PIR capacity for different classes of graphs and compare them with existing results in Table~\ref{tab:pir_graphs_summary}. Our results for multigraph-based systems are presented in Table~\ref{tab:pir_multigraphs_summary}.

\begin{table}[h!]
    \centering
    \caption{Summary of results on the PIR capacity of graphs}
    \begin{tabular}{@{}cccc@{}}
        \toprule
        \textbf{Graph Type} & \thead{\textbf{Rate of the Best-Known Scheme}\\{\textbf{(Capacity Lower Bound)}}} & \textbf{Capacity Upper Bound} \\
        \midrule
        Path graph $\bbP_N$ & $\frac{2}{N}$  (Construction \ref{cons: scheme for P_N}) & $\frac{2}{N}$ (Theorem \ref{Thm_P_N}) \\
        \hline
        Star graph $\bbS_{N+1}$ & $\frac{1}{2\sqrt{N}+1}$ \cite[Theorem 18]{SGT23} & $\frac{1}{\sqrt{2N}-\frac{1}{2}}$ (Theorem \ref{Thm_upp_bound_K_M,N}) \\
        \hline
        \thead{Complete bipartite \\ graph $\bbK_{M,N}$} & $\frac{1}{2M\sqrt{N}+M}$ (Theorem \ref{Thm_upp_bound_K_M,N}) & $\frac{1}{\sqrt{2MN}-\frac{M}{2}}$ (Theorem \ref{Thm_upp_bound_K_M,N}) \\
        \hline
        Complete graph $\bbK_N$ & $\frac{6}{5-2^{3-N}}\cdot\frac{1}{N}$ (Construction \ref{cons: scheme for K_N}) & $\frac{2}{N+1}$ \cite[Theorem 9]{SGT23} \\
        \hline
        General graph & The same as $\bbK_N$ & $\min\left\{\frac{\Delta}{|E(G)|},\frac{1}{\nu(G)}\right\}$\cite[Theorem 1]{SGT23} \\
        \bottomrule
    \end{tabular}
    \label{tab:pir_graphs_summary}
\end{table}

\begin{table}[h!]
    \centering
    \caption{Summary of results on the PIR capacity of $r$-multigraphs}
    \begin{tabular}{@{}ccccc@{}}
        \toprule
        \textbf{Graph Type} & \thead{\textbf{Rate of the Best-Known Scheme}\\{\textbf{(Capacity Lower Bound)}}} & \textbf{Capacity Upper Bound} \\
        \midrule
        Multi-path $\bbP_{N}^{(r)}$ & $\frac{2}{N}\cdot \left(2-\frac{1}{2^{r-1}}\right)^{-1}$ (Theorem \ref{Thm: achievable_lbnd}) & $\begin{cases}
            \frac{2}{N}\cdot \left(2-\frac{1}{2^{r-1}}\right)^{-1},~\text{$N$ even};\\
            \frac{2}{N-1}\cdot \left(2-\frac{1}{2^{r-1}}\right)^{-1},~\text{$N$ odd}.
        \end{cases}$ (Theorem \ref{Thm: Capacity upper bound for multigraphs}) \\
        \hline
        Multi-cycle $\bbC_{N}^{(r)}$ & $\frac{2}{N+1}\cdot \left(2-\frac{1}{2^{r-1}}\right)^{-1}$ (Theorem \ref{Thm: achievable_lbnd}) & $\begin{cases}
            \frac{2}{N}\cdot \left(2-\frac{1}{2^{r-1}}\right)^{-1},~\text{$r\geq 2$ (Theorem \ref{Thm: Capacity upper bound for multigraphs})};\\
            \frac{2}{N+1},~\text{$r=1$~\cite[Theorem 9]{SGT23}}.
        \end{cases}$ \\
        \hline
        Multi-star $\bbS_{N+1}^{(r)}$ & $\frac{2}{N+1}\cdot \left(2-\frac{1}{2^{r-1}}\right)^{-1}$ (Theorem \ref{Thm: achievable_lbnd}) & $\begin{cases}
            \left(2-\frac{1}{2^{r-1}}\right)^{-1},~\text{$r\geq 2$ (Theorem \ref{Thm: Capacity upper bound for multigraphs})};\\
            \frac{1}{\sqrt{2N}-\frac{1}{2}},~\text{$r=1$ (Theorem \ref{Thm_upp_bound_K_M,N})}.
        \end{cases}$ \\
        \hline
        Complete multigraph $\bbK_N^{(r)}$ & $\frac{6}{5-2^{3-N}}\cdot\frac{1}{N}\cdot \left(2-\frac{1}{2^{r-1}}\right)^{-1}$ (Theorem \ref{Thm: achievable_lbnd}) & $\frac{1}{N-(N-1)2^{-r}}$ (Theorem \ref{Thm: multigraph extension of Thm9 in Zachi's paper}) \\
        \hline
        General $r$-multigraph & The same as $\bbK_N^{(r)}$ & $\min\left\{\frac{\Delta}{|E(G)|},\frac{1}{\nu(G)}\right\}\cdot \left(2-\frac{1}{2^{r-1}}\right)^{-1}$ (Theorem \ref{Thm: Capacity upper bound for multigraphs}) \\
        \bottomrule
    \end{tabular}
    \label{tab:pir_multigraphs_summary}
\end{table}

The rest of the paper is organized as follows. In Section \ref{sec: preliminaries}, we formally define our problem setting and present preliminary results on the relationships between PIR capacities of graphs and their subgraphs. Section \ref{sec: PIR over graphs} focuses on establishing PIR capacity bounds for specific classes of graphs. Then, in Section \ref{sec: PIR over multigraphs}, we extend our analysis to derive PIR capacity bounds for general $r$-multigraphs. Finally, Section \ref{sec: conclusion} concludes the paper by highlighting some open problems for future research.

\section{Preliminaries}\label{sec: preliminaries}

Throughout the paper, we use the following standard notations. For integers $1 \leq m \leq n$, let $[m:n] \triangleq \{m, m+1, \ldots, n\}$ and $[n] \triangleq [1:n]$. For a vector $\bbx = (x_1, x_2, \ldots, x_n)$ and a subset $R \subseteq [n]$, let $\bbx|_R$ denote the vector obtained by projecting the coordinates of $\bbx$ onto $R$. If $R = [m_1: m_2]$ for some $1 \leq m_1 \leq m_2 \leq n$, we write $\bbx|_R$ as $\bbx[m_1: m_2]$ for convenience. For a subset $A$ of $[n]$ and a positive integer $t$, we use $\binom{A}{\leq t}$ to denote the family of all subsets of $A$ with size at most $t$, and we use $2^{A}$ to denote the family of all subsets of $A$. We use $H(\cdot)$ to denote the binary entropy function, and $I(\cdot\,;\,\cdot)$ to denote the mutual information between two random variables. Finally, we use $G = (V, E)$ to denote the simple graph with a finite vertex set $V$ and edge set $E \subseteq \binom{V}{2}$. 

\subsection{Problem Setting}\label{subsec: problem setting}

Let $\cS=\{S_1,S_2,\ldots,S_N\}$ denote $N$ non-colluding servers and $\cW=\{W_1,W_2,\ldots,W_K\}$ denote $K$ independent files. Each $W_{i}\in \Ftwo^{L}$ is a binary vector of length $L$ chosen uniformly at random from all the vectors of $\Ftwo^{L}$, hence,
\begin{align}
    H(W_1,\ldots,W_K)&=H(W_1)+\cdots+H(W_K)=K\cdot H(W_1)=KL.
    \label{eq1_problem_setting}
\end{align}

In the PIR problem, a user privately generates $\theta\in [K]$ and wishes to retrieve $W_{\theta}$ while keeping $\theta$ hidden from each server. Let $\cQ\triangleq\{Q_i=Q_i(\theta):i\in [N]\}$ denote the set of all queries generated by the user. Since the user has no information on the content of the files, the queries are independent of them, which means that
\begin{align}\label{eq2_problem_setting}
    I(W_1,\ldots,W_K;Q_1,\ldots,Q_N)=0.
\end{align}
Upon receiving its query $Q_i$ from the user, server $S_i$ generates and replies with an answer $A_i=A_i(\theta)$. We denote $\cW_{i}$ as the set of files stored at $S_{i}$. Then, $A_i$ is a function of $Q_i$ and $\cW_{i}$, which implies that 
\begin{align}\label{eq3_problem_setting}
    H(A_i|Q_{i},\cW_{i})=0,~\forall~i\in [N].
\end{align}

A PIR scheme has two basic requirements: \emph{reliability} and \emph{privacy}, formally described as follows.

\textbf{Reliability}: The user should be able to retrieve the desired file $W_{\theta}$ from the received answers $A_i$ with zero probability of error, hence, 
    \begin{align}\label{eq4_problem_setting}
        H(W_\theta|A_{[N]})=0.
    \end{align}

\textbf{Privacy}: Each server learns no information about the desired file index $\theta$. That is, for any $i\in [N]$ and $\theta\in [K]$, it holds that
     \begin{align}\label{eq5_problem_setting}
        (Q_i(1),A_{i}(1),\cW_{i})\sim (Q_i(\theta),A_{i}(\theta),\cW_{i}),
    \end{align}
    where we use $X \sim Y$ to indicate that the random variables $X$ and $Y$ have the same distribution. Moreover, when $\theta$ is chosen uniformly at random from $[K]$, this is equivalent to
    \begin{align}\label{eq5.1_problem_setting}
        H(\theta|Q_{i},\cW_{i})=H(\theta)=\log(K).
    \end{align}

We define the PIR rate of a retrieval scheme $ T $ as the ratio between the retrieved file size in bits, and the total number of bits downloaded as answers, i.e.,
\begin{align}\label{eq_rate}
    R_{T,L} = \frac{L}{\sum_{i \in [N]} H(A_i)}.
\end{align}
In the sequel, we will omit the subscripts $ L $ and $ T $ for simplicity. 

In this paper, we limit our scope to the PIR problem over \emph{graph-based replication systems}. In such a system, vertices represent servers and (hyper-)edges represent files. A (hyper-)edge is incident to a vertex if a copy of the corresponding file is stored on the corresponding server. Additionally, we also allow the underlying graph to contain multiple (hyper-)edges, and we refer to such systems as \emph{multigraph-based replication systems}. Moreover, a graph/multigraph-based replication system is called an $r$-replication system for some positive integer $r$, if each file is replicated $r$ times across $r$ distinct servers. For convenience, we say that $T$ is a PIR scheme for a graph/multigraph $G$, if $T$ is a retrieval scheme for the PIR problem over the graph-based replication system corresponding to $G$.

For a $2$-replication system with server sets $\cS$ and file sets $\cW$, we use $G = (\cS, \cW)$ to denote the underlying graph/multigraph with the vertex set $\cS$, where a file $W_k \in \cW$ is viewed as the edge $\{S_i, S_j\}$ if the file is stored on servers $S_i$ and $S_j$. The PIR capacity of $ G $ is defined as the best possible rate for arbitrarily large file sizes, i.e.,
\begin{align}\label{eq_capacity}
    \mathscr{C}(G) = \sup_{T,L} R_{T,L}.
\end{align}
Moreover, we say that a PIR scheme for $G$ is \emph{optimal} if its rate equals the capacity $\mathscr{C}(G)$.

By the privacy requirement, one can easily obtain the following result, which will be useful in proving our PIR capacity upper bounds.

\begin{proposition}\label{Prop2 in Zachi's paper}\cite[Proposition 2]{SGT23}
    Given a set of file indices $J \subseteq [K]$, we denote $W_{J} \triangleq \{W_{i} : i \in J\}$. Then, for any answer $A_i$, $i \in [N]$, any requested file index $k \in [K]$, and any $J \subseteq [K]$,
    \begin{align}
        H(A_i | Q_i, W_{J}, \theta = k) = H(A_i | Q_i, W_{J}).
    \end{align}
\end{proposition}

\subsection{PIR Reductions by Graph Decomposition}\label{subsec: PIR reduction}

In this section, we establish a result that relates the PIR capacity of a graph to the PIR capacities of its subgraphs. This result is useful for constructing PIR schemes for specific graph classes in subsequent sections. 

Before presenting the result, we introduce the following definitions. Let $H_1 = (V_1, E_1)$ and $H_2 = (V_2, E_2)$ be two graphs. We say that a graph $G = (V, E)$ is a \emph{vertex-disjoint union} of $H_1$ and $H_2$ if $V = V_1 \cup V_2$, $E = E_1 \cup E_2$, and $V_1 \cap V_2 = \emptyset$. Similarly, $G$ is called an \emph{edge-disjoint union} of $H_1$ and $H_2$ if $V = V_1 \cup V_2$, $E = E_1 \cup E_2$, and $E_1 \cap E_2 = \emptyset$. Clearly, if $G$ is a vertex-disjoint union of $H_1$ and $H_2$, then it is also an edge-disjoint union of $H_1$ and $H_2$.

Consider the PIR problem for a graph $G$ that is an edge-disjoint union of two subgraphs, $H_1$ and $H_2$. Suppose the user desires a file in $H_1$. Then, one can simply run a PIR scheme $T_1$ for $H_1$ to retrieve the file. However, doing so would result in servers receiving queries only related to files in $H_1$, thereby revealing to the servers storing files in $H_2$ that the desired file lies in $H_1$. To preserve privacy, the user can simultaneously send queries regarding files in $H_2$---according to a PIR scheme $T_2$ for $H_2$---while executing $T_1$. These additional queries are used to retrieve a random file in $H_2$. Since $H_1$ and $H_2$ are edge-disjoint, the queries sent to each server according to $T_1$ and $T_2$ are independent. This results in a PIR scheme for $G$ in which reliability is ensured by $T_1$, and privacy is maintained jointly by $T_1$ and $T_2$. We refer to this as a scheme for $G$ obtained by independently applying $T_1$ and $T_2$.

The following result implies that an optimal PIR scheme for a graph $G$ can be obtained by independently applying optimal schemes to each of its connected components, i.e., the set of vertices that are reachable from every other vertex via a path. Consequently, the task of designing optimal schemes for general graphs reduces to separately designing optimal schemes for connected graphs. 

\begin{theorem}\label{thm-stam}
    If $G$ is an edge-disjoint union of $H_1$ and $H_2$, then
    \begin{align}
    \label{stam}
      \mathscr{C}(G) \geq \left(\mathscr{C}(H_1)^{-1}+\mathscr{C}(H_2)^{-1}\right)^{-1}.
    \end{align}
Furthermore, equality holds in \eqref{stam} if $H_1$ and $H_2$ are also vertex-disjoint.
\end{theorem}

Since the proof of Theorem~\ref{thm-stam} is routine and straightforward, we refer the reader to Appendix~\ref{appendix: PIR reduction}.

\section{Improved Capacity Bounds and Schemes for PIR over Graphs}\label{sec: PIR over graphs}

In this section, we study the PIR capacity of several graph classes. First, we determine the PIR capacity of path graphs by establishing an upper bound and constructing an explicit scheme that achieves it. Next, we derive improved upper and lower bounds on the PIR capacity of complete bipartite graphs. Finally, we present a scheme for complete graphs that attains capacity for $N=3$ and achieves a higher rate than all existing schemes for $N \geq 4$.

\subsection{The PIR Capacity of Path Graphs}\label{subsec: PIR over P_N}

For $N \geq 2$, let $\bbP_N$ denote the path graph on $N$ vertices, labeled $S_1, \dots, S_N$. In the replication system based on $\bbP_N$, we assume without loss of generality that server $S_1$ stores file $W_1$, each intermediate server $S_i$ for $i \in [2:N-1]$ stores files $W_{i-1}$ and $W_i$, and server $S_N$ stores file $W_{N-1}$. See Figure \ref{fig1:pir-path-graph} for an illustration of the case $N=3$.
\begin{figure}[h]
    \centering
        \begin{tikzpicture}[
            every node/.style={circle, draw, inner sep=0.3pt, minimum size=0.15cm}, 
            every edge/.style={draw, thick}, 
            edge label/.style={draw=none, rectangle, inner sep=2pt, font=\small}
        ]
        \node (A) at (0, 0) {$S_1$};
        \node (B) at (1.5, 0) {$S_2$};
        \node (C) at (3, 0) {$S_3$};

        \draw[thick] (A) to node[edge label, below] {$W_{1}$} (B);

        \draw[thick] (B) to node[edge label, below] {$W_{2}$} (C);
        \end{tikzpicture}
        \caption{The replication system based on $\bbP_3$}
        \label{fig1:pir-path-graph}
\end{figure}

The following theorem establishes the exact PIR capacity of $\bbP_N$.

\begin{theorem}\label{Thm_P_N}
For positive integer $N\geq 2$, the path graph $\bbP_{N}$ has PIR capacity $\mathscr{C}(\bbP_{N})= \frac{2}{N}$.
\end{theorem}

The proof of Theorem~\ref{Thm_P_N} consists of two parts: an upper bound on the capacity, and a matching lower bound achieved via a PIR scheme construction. We begin with the proof of the capacity upper bound.

To prove the capacity upper bound $\mathscr{C}(\bbP_N)\leq \frac{2}{N}$, we need the following results from \cite{SGT23}.

\begin{lemma}\label{Lem3 in Zachi's paper}\cite[Lemma 3]{SGT23}
    Let $S_i,S_j\in [N]$ be two distinct servers that share the $k$-th file $W_k\in \cW$, then
    \begin{align}
        L & \leq H(A_i|\cW\setminus\{W_k\},\cQ)+H(A_j|\cW\setminus\{W_k\},\cQ)\\
          & \leq H(A_i)+H(A_j).
    \end{align}
\end{lemma}

\begin{theorem}\label{Thm6 in Zachi's paper}\cite[Theorem 6]{SGT23}
    Given a PIR scheme for a graph $G = (\cS, \cW)$, then for any server $S_{i} \in \cS$ with degree $\delta$ and neighbor set $N(S_{i}) = \{S_{i_1}, \ldots, S_{i_\delta}\}$, the following holds:
    \begin{align}
    H(A_{i}|\cQ) \geq \sum_{j=1}^{\delta} \max \left\{ 0, L - \sum_{\ell=j}^{\delta} H(A_{i_\ell}|\cQ) \right\}.
    \end{align}
\end{theorem}

\begin{IEEEproof}[Proof of Upper Bound in Theorem \ref{Thm_P_N}]
    Let $T$ be any feasible PIR scheme for the path graph $\bbP_N$ with rate $R$ and file length $L$. It suffices to show that $R \leq \frac{2}{N}$. We proceed by analyzing the cases based on the parity of $N$.

    When $N$ is even, assume $N = 2t$ for some positive integer $t$. Then, we have
    \begin{align}
        \sum_{i=1}^{N} H(A_i) &= \sum_{j=1}^{t} \left(H(A_{2j-1}) + H(A_{2j})\right) \\
        &\geq tL, \label{eq1_pf_upp_bound_P_N}
    \end{align}
    where \eqref{eq1_pf_upp_bound_P_N} follows from Lemma \ref{Lem3 in Zachi's paper} and the fact that servers $S_{2j-1}$ and $S_{2j}$ share the file $W_{2j-1}$ for each $1 \leq j \leq t$. Then, it follows that
    \begin{align}
        R = \frac{L}{\sum_{i=1}^{N} H(A_i)} \leq \frac{1}{t} = \frac{2}{N}.
    \end{align}

    When $N$ is odd, assume $N=2t+1$ for some positive integer $t$. Similarly, by Lemma \ref{Lem3 in Zachi's paper} and the fact that servers $S_{2j}$ and $S_{2j+1}$ share the file $W_{2j}$, we have
    \begin{align}
        \sum_{i=1}^{N} H(A_i) &= H(A_1) + H(A_2) + H(A_3) + \sum_{j=2}^{t} \left(H(A_{2j}) + H(A_{2j+1})\right) \\
        &\geq H(A_1) + H(A_2) + H(A_3) + (t-1)L. \label{eq2_pf_upp_bound_P_N}
    \end{align}
    We claim that $H(A_1) + H(A_2) + H(A_3) \geq \frac{3L}{2}$. Then, by \eqref{eq2_pf_upp_bound_P_N}, it holds that
    \begin{align}
    \sum_{i=1}^{N} H(A_i) \geq \left(\frac{2t+1}{2}\right) L = \frac{NL}{2},
    \end{align}
    which implies $R \leq \frac{2}{N}$.

    Now, we proceed to show that $H(A_1) + H(A_2) + H(A_3) \geq \frac{3L}{2}$. Consider the following two cases:
    \begin{itemize}
        \item If $H(A_1) \geq \frac{L}{2}$, then, since servers $S_2$ and $S_3$ share the file $W_2$, the claim follows directly by Lemma \ref{Lem3 in Zachi's paper}.
        \item If $H(A_1) < \frac{L}{2}$, note that server $S_2$ has the neighbor set $\{S_1,S_3\}$. Thus, by Theorem \ref{Thm6 in Zachi's paper} and the fact that $H(A_i) \leq L$ for every $i \in [N]$, we have
        \begin{align}
            H(A_2) &\geq L - H(A_1) + \max\left\{0, L - H(A_1) - H(A_3)\right\} \\
            &\geq 
            \begin{cases}
                2L - 2H(A_1) - H(A_3), &\text{if } H(A_1) + H(A_3) < L; \\
                L - H(A_1), &\text{otherwise}.
            \end{cases} \label{eq3_pf_upp_bound_P_N}
        \end{align}
        
        When $H(A_1) + H(A_3) < L$, using \eqref{eq3_pf_upp_bound_P_N} and the assumption that $H(A_1) < \frac{L}{2}$, we get
        \begin{align}
            H(A_1) + H(A_2) + H(A_3) &\geq 2L - H(A_1) \\
            &> \frac{3L}{2}.
        \end{align}
        Otherwise, we have $H(A_1) + H(A_3) \geq L$. Then, by \eqref{eq3_pf_upp_bound_P_N} and the assumption that $H(A_1) < \frac{L}{2}$, we get
        \begin{align}
            H(A_1) + H(A_2) + H(A_3) &\geq L + H(A_2) \\
            &\geq 2L - H(A_1) > \frac{3L}{2}.
        \end{align}
    \end{itemize}
    Therefore, $H(A_1) + H(A_2) + H(A_3) \geq \frac{3L}{2}$ holds in both cases. This concludes the proof.
\end{IEEEproof}

Next, we present a scheme construction for $\bbP_N$ that achieves the capacity upper bound. The following example provides an illustration of the scheme for the case when $N = 3$.

\begin{example}\label{ex_PIR for P3}
The replication system based on $\bbP_3$ consists of three servers, $S_1$, $S_2$, and $S_3$, and two files, $\{W_1, W_2\}$. As illustrated in Figure~\ref{fig1:pir-path-graph}, $S_1$ stores $W_1$, $S_3$ stores $W_2$, and $S_2$ stores both $W_1$ and $W_2$. 

Suppose that each file consists of two bits, and the user wants to privately retrieve $W_{\theta}$ for some $\theta \in [2]$. To do so, the user first picks a permutation of $[2]$ uniformly at random and applies it to the bit indices of file $W_i$, for each $i \in [2]$. Denote the resulting permuted file as $w_i$, $i \in [2]$, and write $w_1 = (a_1, a_2)$ and $w_2 = (b_1, b_2)$. The user then sends queries to retrieve one answer bit from each of $S_1$, $S_2$, and $S_3$, according to Table~\ref{tab:base_answers}.
\begin{table}[h]
\centering
\begin{tabular}{|c|c|c|c|}
    \hline
     & $S_1$ & $S_2$ & $S_3$\\
    \hline
    $\theta=1$ & $a_1$ & $a_2+b_2$ & $b_2$\\
    \hline
    $\theta = 2$ & $a_1$ & $a_1+b_1$ & $b_2$\\
    \hline
\end{tabular}
\vspace{0.2cm}
\caption{Answer table for $\bbP_3$.}
\label{tab:base_answers}
\end{table}

Clearly, the user can decode the desired file for both cases $\theta = 1$ and $\theta = 2$, and the rate of the scheme is $\frac{2}{3}$. Also, since the bit indices of each file are permuted uniformly at random, the queries appear uniformly distributed regardless of the value of $\theta$. This guarantees privacy.
\end{example}

\begin{construction}[A capacity-achieving PIR Scheme for $\bbP_N$]\label{cons: scheme for P_N}
    Suppose that for each $j\in [N-1]$, the file $W_j=(W_{j}(1),W_{j}(2))\in \Ftwo^{2}$ is a binary vector of length 2. The retrieval scheme is described as follows:
    
    \begin{itemize}
        \item [(a)] The user chooses a file index $\theta \in [N-1]$ and $N-1$ permutations $\sigma_j: [2] \rightarrow [2]$, independently and uniformly at random, from the set of all permutations on $[2]$. For each $j \in [N-1]$, apply the permutation $\sigma_j$ to the bits of $W_j$, and denote the resulting permuted file as:
        \begin{align}
            w_j \triangleq \left( W_j(\sigma_j(1)), W_j(\sigma_j(2))\right) = \left(w_{j}(1), w_j(2)\right).
        \end{align}
        
        \item [(b)] The user sends the query $Q_i$ to each server $S_i,~i \in [N]$, defined as:
        \begin{align}
           Q_i = \begin{cases}
            \sigma_1(1),  & \text{if } i=1;\\
            (\sigma_{i-1}(1), \sigma_i(1)), & \text{if } 2\leq i\leq \theta; \\
            (\sigma_{i-1}(2), \sigma_i(2)) , & \text{if } \theta<i\leq N-1;\\
            \sigma_{N-1}(2),  & \text{if } i=N.
        \end{cases} 
        \end{align}
        \item [(c)] Each server $S_i,~i \in [N]$ returns the answer $A_i(Q_i)$, defined as:
        \begin{align}
            A_i(Q_i) = 
            \begin{cases}
            w_1(1), & \text{if } i=1;\\
            w_{i-1}(1) + w_i(1), & \text{if } 2\leq i\leq \theta; \\
            w_{i-1}(2) + w_i(2), & \text{if } \theta<i\leq N-1; \\
            w_{N-1}(2), & \text{if } i=N.
            \end{cases}
        \end{align}
    \end{itemize}
\end{construction}

\begin{IEEEproof}[Proof of Lower Bound in Theorem \ref{Thm_P_N}]
    It suffices to show that the scheme given in Construction \ref{cons: scheme for P_N} satisfies the privacy and reliability requirements and has the claimed rate.

    \textbf{Privacy:} Note that $\{\sigma_{j}\}_{j\in [N-1]}$ are chosen independently and uniformly at random. Thus, by its definition, the query $Q_i$ sent to server $S_i$ is uniformly distributed over $[2] \times [2]$ for $i \in [2:N-1]$, and over $[2]$ for $i \in \{1, N\}$. This implies that $Q_i$ reveals no information about $\theta$, which guarantees privacy.
    
    \textbf{Reliability:} By summing all the answers $A_i(Q_i)$ for $1 \leq i \leq \theta$, the user can retrieve $w_{\theta}(1)$, as
    \begin{align}
        w_{\theta}(1) = w_1(1) + \sum_{i=2}^{\theta} \left( w_{i-1}(1) + w_i(1) \right).
    \end{align}
    Similarly, by summing the remaining answers $A_i(Q_i)$ for $\theta + 1 \leq i \leq N$, the user can retrieve $w_{\theta}(2)$, as
    \begin{align}
    w_{\theta}(2) =  \sum_{i=\theta+1}^{N-1} \left( w_{i-1}(2) + w_i(2) \right) + w_{N-1}(2).
    \end{align}
    Thus, the user can recover $W_{\theta}$ using the permutation $\sigma_{\theta}$, which confirms the reliability requirement.

    \textbf{Rate:} The size of each file is 2. Since each server returns a single bit, the rate of the scheme is $\frac{2}{N}$.
\end{IEEEproof}

\subsection{Improved Capacity Bounds for Complete Bipartite Graphs}\label{subsec: PIR over K_M,N}

Let $\bbK_{M,N}$ denote the complete bipartite graph with two sets of vertices of sizes $M$ and $N$ $(M \leq N)$, and let $\bbK_{N}$ denote the complete graph on $N$ vertices. In \cite[Theorem 1]{SGT23}, Sadeh, Gu and Tamo established an upper bound on the PIR capacity of general graphs (see also in Table \ref{tab:pir_graphs_summary}). As a consequence, they showed that the capacity of $\bbK_{M,N}$ satisfies 
\begin{align}
\frac{1}{1-2^{1-(M+N)}}\cdot\frac{1}{M+N}\leq \mathscr{C}(\bbK_{M+N}) \leq \mathscr{C}(\bbK_{M,N}) \leq \frac{1}{M}. \label{previous bd on K_M,N}     
\end{align}
In this subsection, we prove the following bounds on $\mathscr{C}(\bbK_{M,N})$, which improve both the upper and lower bounds compared to those in \eqref{previous bd on K_M,N}.

\begin{theorem}\label{Thm_upp_bound_K_M,N}
The PIR capacity of the complete bipartite graph $\bbK_{M,N}$ satisfies,
\begin{align}\label{eq_PIR capacity of K_M,N}
    \frac{1}{2M\sqrt{N}+M}\leq \mathscr{C}(\bbK_{M,N}) \leq \frac{1}{\sqrt{2MN}-\frac{M}{2}}.
\end{align}
\end{theorem}

\begin{remark}\label{rmk1_upp_bound_K_M,N}
    Note that when $N \geq \frac{9M}{8}$, it holds that
    \begin{align}
        \frac{1}{\sqrt{2MN}-\frac{M}{2}}&\leq \frac{1}{\sqrt{\frac{9M^2}{4}}-\frac{M}{2}}=\frac{1}{M}.
    \end{align}
    Thus, compared to the upper bound in \eqref{previous bd on K_M,N} by Sadeh, Gu and Tamo \cite{SGT23}, the upper bound in Theorem \ref{Thm_upp_bound_K_M,N} is tighter when $N \geq \frac{9M}{8}$. When $\sqrt{N}\geq 5M$, it holds that     
    \begin{align}
        \frac{1}{2M\sqrt{N}+M} &\geq \frac{1}{\frac{2}{5}N+M}\geq\frac{2}{M+N} \\
        &\geq \frac{1}{1-2^{1-(M+N)}}\cdot\frac{1}{M+N}.
    \end{align}
    Thus, the lower bound in Theorem \ref{Thm_upp_bound_K_M,N} is tighter compared to that in \eqref{previous bd on K_M,N} when $N \geq 25M^2$.

    Furthermore, when $M = 1$, the bound in Theorem \ref{Thm_upp_bound_K_M,N} reduces to 
    \begin{align}
        \frac{1}{2\sqrt{N}+1}\leq \mathscr{C}(\bbK_{1,N}) \leq \frac{1}{\sqrt{2N}-\frac{1}{2}}.
    \end{align}
    The lower bound coincides with that of Theorem 18 in \cite{SGT23}. Moreover, since $\sqrt{2N+1} \leq \sqrt{2N} + 1$, the upper bound offers a slight improvement over the upper bound $\mathscr{C}(\bbK_{1,N}) \leq \frac{1}{\sqrt{2N+1}-2}$ given by Theorem 8 in \cite{SGT23}.
\end{remark}

For the proof of the capacity upper bound in Theorem \ref{Thm_upp_bound_K_M,N}, we first introduce some necessary results and definitions.

\begin{itemize}
    \item An \emph{automorphism} of a graph $G = (V, E)$ is a permutation $\pi$ on the vertex set $V$ that maps edges to edges; that is, two vertices $u$ and $v$ form an edge $\{u,v\} \in E$ if and only if $\{\pi(u),\pi(v)\} \in E$. The automorphism group $\mathrm{Aut}(G)$ of a graph $G$ consists of all its automorphisms.
    
    \item A graph $G$ is called \emph{vertex-transitive} if its automorphism group $\mathrm{Aut}(G)$ acts transitively on its vertices, meaning that for any two vertices $u, v \in V(G)$, there exists an automorphism $\pi \in \mathrm{Aut}(G)$ such that $\pi(v) = u$. 
    
    \item A bipartite graph $G = (V, E)$ with vertex bipartition $\{A, B\}$ is called \emph{vertex-part-transitive} if there exists a subgroup $\Gamma$ of $\mathrm{Aut}(G)$ that acts transitively on both $A$ and $B$. This means that for any $a, a' \in A$ and $b, b' \in B$, there exists $\pi \in \Gamma$ such that $\pi(a) = a'$, $\pi(b) = b'$, and $\{a',b'\} \in E$ if and only if $\{a,b\} \in E$.
\end{itemize}

We also require the following variant of Lemma 11 in \cite{SGT23} for vertex-part-transitive bipartite graphs. For the reader's convenience, we defer its proof to the Appendix \ref{pf of variant of Lem11 in Zachi's paper}.

\begin{lemma}\label{Variant of Lem11 in Zachi's paper}
    Let $G$ be a bipartite graph with vertex bipartition $\{U, V\}$, and suppose $G$ is vertex-part-transitive. Then, for any achievable rate $R$, there exists a PIR scheme for $G$ with rate $R$ that satisfies:
    \begin{align}
        H(A_i|\cQ) = H(A_{i'}|\cQ), & \quad \forall ~S_i, S_{i'} \in U; \\
        H(A_j|\cQ) = H(A_{j'}|\cQ), & \quad \forall ~S_j, S_{j'} \in V.
    \end{align}
\end{lemma}

\begin{IEEEproof}[Proof of Theorem \ref{Thm_upp_bound_K_M,N}]
    We start with the proof of the lower bound. Note that $\bbK_{M,N}$ can be viewed as the edge-disjoint union of $M$ copies of the star graph $\bbK_{1,N}$ (also denoted by $\bbS_{N+1}$) over $N+1$ vertices. Recall that the scheme for $\bbK_{1,N}$ in \cite{SGT23} achieves a rate of $\frac{1}{2\sqrt{N}+1}$. Thus, the lower bound in Theorem \ref{Thm_upp_bound_K_M,N} follows directly by Theorem \ref{thm-stam}. 

    Next, we prove the upper bound.

    Consider a PIR scheme for $\bbK_{M,N}$ with vertex bipartition $U$ and $V$. Let $S_1,\ldots,S_{N}$ be the vertices (servers) in $U$, and $S_{N+1},\ldots,S_{N+M}$ be the vertices (servers) in $V$. Since $\bbK_{M,N}$ is vertex-part-transitive, by Lemma \ref{Variant of Lem11 in Zachi's paper}, we have $H(A_i|\cQ) = H(A_{i'}|\cQ)$, for any $i, i' \in [N]$ and $H(A_j|\cQ) = H(A_{j'}|\cQ)$, for any $j, j' \in [N+1:N+M]$.
    
    Let $t$ be the largest integer in $[N]$ such that $H(A_{N-t+1}|\cQ)\leq \frac{L}{t}$, and note that $t\geq 1$, since $H(A_{N}|\cQ)\leq L$. Clearly, by the definition of $t$, it holds that $H(A_{N-t}|\cQ)> \frac{L}{t+1}$. Moreover, since $H(A_i|\cQ) = H(A_{i'}|\cQ)$ for any $i,i' \in [N]$, we can assume that 
    \begin{equation}\label{eq0_pf_upp_bound_K_M,N}
        H(A_{1}|\cQ)=H(A_{2}|\cQ)=\cdots=H(A_{N}|\cQ)=\delta L
    \end{equation}
    for some $\frac{1}{t+1}<\delta\leq \frac{1}{t}$. Then, for every $j\in [N+1:N+M]$, by Theorem \ref{Thm6 in Zachi's paper}, we have
    \begin{align}
        H(A_j|\cQ)&\geq \sum_{i=1}^{N} \max \left\{ 0, L - \sum_{\ell=i}^{N} H(A_{\ell}|\cQ) \right\}\\
        &\geq \sum_{i=N-t+1}^{N} \max \left\{ 0, L - \sum_{\ell=i}^{N} H(A_{\ell}|\cQ) \right\} \\
        &= \sum_{i=N-t+1}^{N} \max \left\{ 0, L - (N-i+1)\cdot  \delta L \right\} \label{eq1_pf_upp_bound_K_M,N}\\
        &= tL - \frac{t(t+1)}{2}\cdot \delta L \label{eq2_pf_upp_bound_K_M,N}
    \end{align}
    where \eqref{eq1_pf_upp_bound_K_M,N} follows by \eqref{eq0_pf_upp_bound_K_M,N}, and \eqref{eq2_pf_upp_bound_K_M,N} follows by $\delta\leq \frac{1}{t}$. Thus, \eqref{eq2_pf_upp_bound_K_M,N} implies that
    \begin{align}
        \sum_{i=1}^{N+M} H(A_i) &\geq \sum_{i=1}^{N+M} H(A_i | \cQ) 
         \\
        &= M \cdot H(A_{N+1} | \cQ) + N \cdot H(A_N | \cQ) 
        \\
        &\geq M \left(t L - \frac{t(t+1)}{2} \cdot \delta L \right) + N \delta L 
         \\
        &= \frac{M \delta L}{2} \left(\frac{2t}{\delta} - t^2-t \right) + N \delta L  \\
        &= -\frac{M \delta L}{2} \left(t - \frac{1}{\delta} + \frac{1}{2} \right)^2 + \frac{M \delta L}{2} \left(\frac{1}{\delta} - \frac{1}{2} \right)^2 + N \delta L  \\ 
        &\geq -\frac{M \delta L}{2} \cdot \frac{1}{4} + \frac{M \delta L}{2} \left(\frac{1}{\delta^2} + \frac{1}{4} - \frac{1}{\delta} \right) + N \delta L 
        \label{eq4_pf_upp_bound_K_M,N} \\
        &= \frac{M L}{2 \delta} + N \delta L - \frac{M L}{2} 
       \\
        &\geq \left(\sqrt{2 M N} - \frac{M}{2}\right) L,
        \label{eq5_pf_upp_bound_K_M,N} 
    \end{align}
    where \eqref{eq4_pf_upp_bound_K_M,N} follows since the function $f(x) = -\left(x - \frac{1}{\delta} + \frac{1}{2}\right)^2$ attains its minimum at the boundaries of the interval and given $t \leq \frac{1}{\delta} < t+1$, we have $t - \frac{1}{\delta} + \frac{1}{2} \in \left(-\frac{1}{2}, \frac{1}{2}\right]$, and \eqref{eq5_pf_upp_bound_K_M,N} follows by applying the AM-GM inequality. Therefore, the rate of the scheme satisfies 
    \begin{align}
    \frac{L}{\sum_{i=1}^{M+N}H(A_i)}\leq \frac{L}{\left(\sqrt{2MN}-\frac{M}{2}\right)L}=\frac{1}{\sqrt{2MN}-\frac{M}{2}},
    \end{align}
    as required.
\end{IEEEproof}

\subsection{Improved PIR Scheme for Complete Graphs}\label{subsec: PIR over K_N}

Let $\bbK_{N}$ denote the complete graph over a vertex set of size $N$. As pointed out in~\cite{graphbased_pir} and~\cite{SGT23}, understanding the PIR capacity of $\bbK_{N}$ is of particular importance, as this graph contains the maximum number of files for a given number of servers in graph-based replication systems. Moreover, since any graph $G$ on $N$ vertices is a subgraph of $\bbK_{N}$, any PIR scheme for $\bbK_{N}$ can be converted into a scheme for $G$. This implies that $\mathscr{C}(\bbK_{N}) \leq \mathscr{C}(G)$, which serves as a general lower bound on the PIR capacity of any graph $G$.

In \cite{SGT23}, the PIR capacity of $\bbK_N$ is shown to be at most $\frac{2}{N+1}$. For the lower bound, it is known from \cite{BU19} that $\mathscr{C}(\bbK_3) = \frac{1}{2}$ and $\mathscr{C}(\bbK_4) \geq \frac{3}{10}$. For general $N$, Sadeh, Gu and Tamo \cite{SGT23} provided a general scheme with rate $\frac{2^{N-1}}{2^{N-1}-1} \cdot \frac{1}{N}$ for any $N \geq 2$, which slightly improves upon the previous scheme with rate $\frac{1}{N}$ and demonstrates that the bound of $\frac{1}{N}$ is not optimal. 

In this subsection, we show that the scheme by Sadeh, Gu and Tamo \cite{SGT23} can be further modified to achieve a scheme with rate $\frac{6}{5-2^{3-N}} \cdot \frac{1}{N}$. As a consequence, this yields another capacity-achieving scheme that can be viewed as a reduced version of the one proposed by Banawan and Ulukus in~\cite{BU19} for $\bbK_3$, while also providing an improvement over their result for $\bbK_4$. 

We begin with the following illustrative example of our PIR scheme for $\bbK_3$ with the optimal rate $\frac{1}{2}$.
\begin{example}\label{ex_PIR for K3}
The replication system over $\bbK_3$ consists of $3$ servers, $ S_1 $, $ S_2 $, $ S_3 $ and $3$ files. For each $\{i,j\}\subseteq [3]$, let $W_{i,j}$ be the file stored on servers $S_i$ and $S_j$, as illustrated in Figure \ref{fig:pir-K_3}. 
\begin{figure}[h]
\centering
\begin{tikzpicture}[
    every node/.style={circle, draw, inner sep=0.3pt, minimum size=0.15cm}, 
    every edge/.style={draw, thick}, 
    edge label/.style={draw=none, rectangle, inner sep=2pt, font=\small}
]
\node (A) at (0, 0) {$S_2$};
\node (B) at (1, 1.25) {$S_1$};
\node (C) at (2, 0) {$S_3$};

\draw[thick] (A) to node[edge label, above left] {$W_{1,2}$} (B);

\draw[thick] (A) to node[edge label, below] {$W_{2,3}$} (C);

\draw[thick] (B) to node[edge label, above right] {$W_{1,3}$} (C);
\end{tikzpicture}
\caption{The replication system based on $\bbK_3$}
\label{fig:pir-K_3}
\end{figure}

Suppose that each file consists of $ L = 6 $ bits, and the user wants to privately retrieve $ W_{\theta} $. To do so, the user privately generates 3 independent permutations of $[L]$, denoted by $ \pi_i $, $ i \in [3] $. For $ i \in [L] $, let 
$a_i \triangleq W_{1,2}(\pi_1(i)),$ 
which represents the $ \pi_1(i) $-th bit of $ W_{1,2} $. Similarly, define $b_i \triangleq W_{1,3}(\pi_2(i))$ as the $ \pi_2(i) $-th bit of $ W_{1,3} $, and $c_i \triangleq W_{2,3}(\pi_3(i))$ as the $ \pi_3(i) $-th bit of $ W_{2,3} $. Then, for every possible $ \theta $, the user queries and downloads four bits from each of the servers $ S_1 $, $ S_2 $, and $ S_3 $, as shown in Table~\ref{tab:answers_K3}. 
\begin{table}[h]
\centering
 \begin{tabular}{|c|c|c|c|}   
 \hline
 & $S_1$ & $S_2$ & $S_3$\\
 \hline
 \multirow{4}{*}{$\theta = (1,2)$} & $a_1$ & $a_3$ & $b_2$\\
 & $b_6$ & $c_5$ & $c_4$\\
 & $a_2+b_2$ & $a_4+c_4$ & $b_5+c_5$\\
 & $a_5+b_5$ & $a_6+c_6$ & $b_6+c_6$\\
 \hline
  \multirow{4}{*}{$\theta = (1,3)$} & $a_6$ & $a_2$ & $b_3$\\
 & $b_1$ & $c_4$ & $c_5$\\
 & $a_2+b_2$ & $a_5+c_5$ & $b_4+c_4$\\
 & $a_5+b_5$ & $a_6+c_6$ & $b_6+c_6$\\
 \hline
  \multirow{4}{*}{$\theta = (2,3)$} & $a_2$ & $a_6$ & $b_5$\\
 & $b_4$ & $c_1$ & $c_3$\\
 & $a_5+b_5$ & $a_2+c_2$ & $b_4+c_4$\\
 & $a_6+b_6$ & $a_5+c_5$ & $b_6+c_6$\\
 \hline
 \end{tabular}
 \vspace{0.2cm}
 \caption{Answer table of our PIR scheme for $\bbK_3$.}
 \label{tab:answers_K3}
\end{table}

Clearly, the user can decode the desired file in all three cases. For example, consider the case where $\theta=(1,2)$. According to the answers in Table~\ref{tab:answers_K3}, the user can retrieve all bits in $W_{1,2}$ as follows: 
\begin{itemize} 
\item[1.] $a_1$ directly from $S_1$, and $a_3$ directly from $S_2$; 
\item[2.] $a_2$ by summing $a_2 + b_2$ from $S_1$ with $b_2$ from $S_3$, and $a_4$ by summing $a_4 + c_4$ from $S_2$ with $c_4$ from $S_3$; 
\item[3.] $a_5$ by summing $a_5 + b_5$ from $S_1$ with $c_5$ from $S_2$ and $b_5 + c_5$ from $S_3$, and $a_6$ by summing $a_6 + c_6$ from $S_2$ with $b_6$ from $S_1$ and $b_6 + c_6$ from $S_3$. 
\end{itemize}  

Furthermore, since the permutation of bits is unknown to each server, the query for each server appears uniformly distributed regardless of the value of $ \theta $. This results in a PIR scheme for $ \bbK_3 $ with rate $ \frac{6}{4 \times 3} = \frac{1}{2} $, which achieves the capacity upper bound $ \frac{2}{N+1} $ for $ N = 3 $.
\end{example}

Now, we present the formal description of our scheme construction.
\begin{construction}[A PIR scheme for $\bbK_N$]\label{cons: scheme for K_N}
    The system for $\bbK_N$ consists of $N$ servers, $ S_i $, $ i\in [N] $, and ${N \choose 2}$ files, $ W_{i,j} $, $ \{i,j\}\in {[N]\choose 2} $, which is uniquely stored on servers $S_i$ and $S_j$. Assume further that each file is a binary vector of length $L = 3\cdot 2^{N-2}$.

    To begin with, the user privately and independently chooses ${N \choose 2}$ permutations of $[L]$ uniformly at random, denoted by $ \pi_{i,j} $, for each $ \{i,j\} \in \binom{[N]}{2} $. We denote 
    \begin{align}
    w_{i,j} = \big(w_{i,j}(1), w_{i,j}(2), \ldots, w_{i,j}(L)\big)
    \end{align}
    with $w_{i,j}(\ell) = W_{i,j}(\pi_{i,j}(\ell))$, as the resulting file after applying permutation $ \pi_{i,j} $ to the bits of $ W_{i,j} $.

    \textbf{Generating Queries}: Assume that $ W_{i,i'} $ is the desired file, and let $ \mathcal{F} $ be the family of subsets of $[N]$ that contain exactly one of $ i $ or $ i' $, i.e.,
    \begin{align}\label{eq1F_cons_K_N}
        \mathcal{F} = \{ P \subseteq [N] : |P \cap \{i, i'\}| = 1 \}.
    \end{align}
    Clearly, $ |\mathcal{F}| = 2^{N-1} $. Let $ \phi : \mathcal{F} \to [2^{N-1}] $ be a fixed bijection. For example, for each $P\in \cF$, we set
    $\phi(P)=\ell$, if $P$ is the $\ell$-th set in $\cF$ under the lexicographic order. Let $\cG$ be the family of all disjoint set pairs $(P_1,P_2)$ satisfying $|P_1|\leq |P_2|$ and $P_1\cup P_2=[N]\setminus \{i,i'\}$.
    Clearly, $|\cG|=2^{N-3}$. Let $ \varphi : \mathcal{G} \to [2^{N-1}+1:2^{N-1}+2^{N-3}] $ be a fixed bijection. For example, for each $(P_1,P_2)\in \cG$, we set
    \begin{align}
        \varphi(P_1,P_2)=\ell + 2^{N-1},
    \end{align}
    if $P_1$ is the $\ell$-th set in ${{[N]\setminus\{i,i'\}}\choose \leq \frac{N-2}{2}}$ under the lexicographic order for some $\ell\in [2^{N-3}]$.
    
    Next, for each server $ S_j $, $ j \in [N] $, we construct a map $ \sigma_j: 2^{N(j)} \setminus \{\emptyset\} \rightarrow [L] $, which is only known to the user, where $N(j)=[N]\setminus \{j\}$ is the neighbor set of $S_j$ in $\bbK_N$.

    For server $ S_i $, let $ P \subseteq N(i) $ be a set that contains $ i' $ and define
    \begin{align}\label{eq2_cons_K_N}
        \sigma_i(P) = \phi(\{i\}\cup (P\setminus\{i'\})).
    \end{align}
    Note that there are exactly $ 2^{N-2} $ sets $ P \subseteq N(i)$ containing $i'$. For $ P\subseteq N(i)\setminus \{i'\}$, assume that $|P|\leq (N-2)/2$, we define
    \begin{align}
        \sigma_i(P) & = \sigma_i(N(i)\setminus (P\cup \{i'\}))  \\
        & = \varphi(P,N(i)\setminus (P\cup \{i'\}))+2^{N-3}. \label{eq3_cons_K_N}
    \end{align}
    Clearly, the $\sigma_i$ defined above assigns each nonempty subset $P\subseteq N(i)$ a unique index in $[L]$. Similarly, for server $S_{i'}$, we define
    \begin{align}\label{eq4_cons_K_N}
        \sigma_{i'}(P) = \phi(\{i'\}\cup (P\setminus\{i\}))
    \end{align}
    for each $ P \subseteq N(i') $ containing $i$, and
    \begin{align}
        \sigma_{i'}(P) & = \sigma_{i'}(N(i')\setminus (P\cup \{i\}))  \\
        & = \varphi(P,N(i')\setminus (P\cup \{i\})).\label{eq5_cons_K_N}
    \end{align}
    for each $ P\subseteq N(i')\setminus \{i\}$ with $|P|\leq (N-2)/2$.
    
    For sever $ S_j $ with $ j\notin \{i, i'\} $, let $ P $ be a subset of $N(j)$ such that $ P\in \cF $, we define
    \begin{align}\label{eq6_cons_K_N}
        \sigma_j(P) = \phi(P\cup \{j\}).
    \end{align}
    Note that there are exactly $ 2^{N-2} $ such sets $ P $. For $P\subseteq N(j)$ such that $\{i,i'\}\subseteq P$, denote $\tilde{P}=\{j\}\cup(P\setminus\{i,i'\})$. Then we define
    \begin{align}\label{eq7_cons_K_N}
        \sigma_j(P) = \begin{cases}
            \varphi(\tilde{P},[N]\setminus (\tilde{P}\cup \{i,i'\})),~\text{$|\tilde{P}|\leq (N-2)/2$};\\
            \varphi([N]\setminus (\tilde{P}\cup \{i,i'\}),\tilde{P}),~\text{otherwise}.
        \end{cases}
    \end{align}
    Note that there are exactly $2^{N-3}$ such sets $P$. For the remaining $2^{N-3} - 1$ non-empty subsets of $N(j)$, each of which contains neither $i$ nor $i'$, we arbitrarily assign an unused index from $[2^{N-1}]$ to each subset, ensuring that $\sigma_j$ remains injective.
    
    Next, we describe the bits that the user downloads from each server, which specify the queries each server receives and the answers it provides.

    \textbf{Answers from Servers}: The user downloads $2^{N-1} - 1 + 2^{N-3}$ bits from each server $S_j$: one bit for each nonempty subset of $N(j)$, totaling $2^{N-1} - 1$ bits, and one bit for each set pair $(P_1, P_2)$ in $\cG$, totaling $2^{N-3}$ bits.

    Each of these bits is from some sum of the files stored on $ S_j $. Specifically, they are calculated as follows. For a nonempty subset $ P \subseteq N(j) $, the corresponding downloaded bit is defined as:
    \begin{align}\label{eq8_cons_K_N}
        D_P^{j} \triangleq \sum_{\ell \in P} w_{j,\ell}\left({\sigma_j(P)}\right).
    \end{align}
    When $j=i$, for a set pair $ (P_1, P_2) $ in $\cG$, the corresponding downloaded bit is defined as:
    \begin{align}\label{eq9_cons_K_N}
        D_{(P_1,P_2)}^{i} \triangleq \sum_{\ell \in N(i)} w_{i,\ell}\left({\varphi(P_1,P_2)}\right).
    \end{align}
    When $j\neq i$, the downloaded bit corresponds to the set pair $ (P_1, P_2) $ in $\cG$ is defined as:
    \begin{align}\label{eq10_cons_K_N}
        D_{(P_1,P_2)}^{j} \triangleq \sum_{\ell \in N(j)} w_{j,\ell}\left({\varphi(P_1,P_2)}+2^{N-3}\right).
    \end{align}
\end{construction}

\begin{remark}\label{rmk: Ex2 continue}
    Taking the scheme for $\bbK_3$ in Example \ref{ex_PIR for K3} as an example, we demonstrate how the user's query bits are generated, specifically the mapping $\sigma_j$ for each $j \in [3]$.

    Recall that in Example \ref{ex_PIR for K3}, $w_{1,2}=(a_1,\ldots,a_6)$, $w_{1,3}=(b_1,\ldots,b_6)$ and $w_{2,3}=(c_1,\ldots,c_6)$. Assume that $W_{1,2}$ is the desired file. Then, by \eqref{eq1F_cons_K_N} and the definition of $\cG$, we have $\cF=\{ \{1\}, \{2\}, \{1,3\}, \{2,3\}\}$ and $\cG=\{ (\emptyset, \{3\} ) \}$. 

    We set $ \phi : \mathcal{F} \to [4] $ as
    \begin{align}
        \phi(\{1\})=1, \phi(\{1,3\})=2, \phi(\{2\})=3, \phi(\{2,3\})=4,
    \end{align}
    and $\varphi: \cG \to \{5\}$ as $\varphi(\emptyset, \{3\})=5$. Thus, by \eqref{eq2_cons_K_N} and \eqref{eq3_cons_K_N}, $\sigma_1$ is defined as
    \begin{align}
        \sigma_1(\{2\}) & =\phi(\{1\})=1, \\
        \sigma_1(\{2,3\}) & =\phi(\{1,3\})=2, \\
        \sigma_1(\{3\}) & = \varphi(\emptyset,\{3\})+1=6.
    \end{align}
    By \eqref{eq4_cons_K_N} and \eqref{eq5_cons_K_N}, $\sigma_2$ is defined as
    \begin{align}
        \sigma_2(\{1\}) & =\phi(\{2\})=3, \\
        \sigma_2(\{1,3\}) & =\phi(\{2,3\})=4, \\
        \sigma_2(\{3\}) & = \varphi(\emptyset,\{3\})=5,
    \end{align}
    and by \eqref{eq6_cons_K_N} and \eqref{eq7_cons_K_N}, $\sigma_3$ is defined as
    \begin{align}
        \sigma_3(\{1\}) & =\phi(\{1,3\})=2, \\
        \sigma_3(\{2\}) & =\phi(\{2,3\})=4, \\
        \sigma_3(\{1,2\}) & = \varphi(\emptyset,\{3\})=5.
    \end{align}
    Therefore, by \eqref{eq8_cons_K_N} and \eqref{eq9_cons_K_N}, the bits downloaded from $S_1$ are
    \begin{align}
        D_{\{2\}}^{1}=w_{1,2}(1)=a_1,~& D_{\{2,3\}}^{1}=w_{1,2}(2)+w_{1,3}(2)=a_2+b_2, \\
        D_{\{3\}}^{1}=w_{1,3}(6)=b_6,~& D_{(\emptyset,\{3\})}^{1}=w_{1,2}(5)+w_{1,3}(5)=a_5+b_5;
    \end{align}
    and by \eqref{eq8_cons_K_N} and \eqref{eq10_cons_K_N}, the bits downloaded from $S_2$ and $S_3$ are
    \begin{align}
        D_{\{1\}}^{2}=w_{1,2}(3)=a_3,~& D_{\{1,3\}}^{2}=w_{1,2}(4)+w_{2,3}(4)=a_4+c_4, \\
        D_{\{3\}}^{2}=w_{2,3}(5)=c_5,~& D_{(\emptyset,\{3\})}^{2}=w_{1,2}(6)+w_{2,3}(6)=a_6+c_6, \\
        D_{\{1\}}^{3}=w_{1,3}(2)=b_2,~& D_{\{1,2\}}^{3}=w_{1,3}(5)+w_{2,3}(5)=b_5+c_5, \\
        D_{\{2\}}^{3}=w_{2,3}(4)=c_4,~& D_{(\emptyset,\{3\})}^{3}=w_{1,3}(6)+w_{2,3}(6)=b_6+c_6.
    \end{align}
    This matches the Table \ref{tab:answers_K3}.
\end{remark}

\begin{theorem}\label{Thm: scheme rate for K_N}
The scheme by Construction \ref{cons: scheme for K_N} is a PIR scheme for $\bbK_N$ with rate  
\begin{align}
    \frac{3\cdot2^{N-2}}{2^{N-1}+2^{N-3} - 1} \cdot \frac{1}{N}=\frac{6}{5-2^{3-N}}\cdot \frac{1}{N}.
\end{align}
\end{theorem}

\begin{IEEEproof}
In the following, we show that the scheme by Construction \ref{cons: scheme for K_N} satisfies the reliability and privacy requirements and achieves the claimed rate.

\textbf{Reliability:}   We start by showing that the user can compute every bit of $ w_{i, i'} $ using the scheme, and therefore retrieve the desired file $ W_{i,i'} $.

For $ t \in [2^{N-1}] $, let $ P \in \cF $ be the unique set such that $ \phi(P) = t $. Assume that $ \{i\} = P \cap \{i, i'\} $. Clearly, it holds that $\{i'\}\cup (P\setminus \{i\})\in \cF \cap 2^{N(i)}$ and $P\setminus\{j\}\in \cF\cap 2^{N(j)}$ for each $j\in [N]\setminus \{i,i'\}$. Then, using the downloaded bits from the servers, the user computes the following bit:  
\begin{align}
     D_{\{i'\}\cup (P\setminus \{i\})}^{i}&+\sum_{j\in P\setminus \{i\}}D_{P\setminus\{j\}}^{j} \nonumber \\
    = & \sum_{\ell \in \{i'\}\cup (P\setminus \{i\})} w_{i,\ell}\left({\sigma_i(\{i'\}\cup (P\setminus \{i\}))}\right)+\sum_{j\in P\setminus \{i\}}\sum_{\ell\in P\setminus\{j\}} w_{j,\ell}\left({\sigma_j(P\setminus\{j\})}\right) \label{eq1: Thm scheme rate for K_N}\\
    = & \sum_{\ell \in \{i'\}\cup (P\setminus \{i\})} w_{i,\ell}\left(\phi(P)\right)+\sum_{j\in P\setminus \{i\}}\sum_{\ell\in P\setminus\{j\}} w_{j,\ell}\left(\phi(P)\right) \label{eq2: Thm scheme rate for K_N}\\
    = & w_{i,i'}(t)+\sum_{\ell \in P\setminus \{i\}} w_{i,\ell}(t)+\sum_{j\in P\setminus \{i\}}\sum_{\ell\in P\setminus \{j\}} w_{j,\ell}(t) \label{eq3: Thm scheme rate for K_N}\\
    = & w_{i,i'}(t), \label{eq4: Thm scheme rate for K_N}
\end{align}
where \eqref{eq1: Thm scheme rate for K_N} follows by \eqref{eq8_cons_K_N}, \eqref{eq2: Thm scheme rate for K_N} follows by \eqref{eq2_cons_K_N} and \eqref{eq6_cons_K_N}, \eqref{eq3: Thm scheme rate for K_N} follows by the assumption that $\phi(P)=t$ and \eqref{eq4: Thm scheme rate for K_N} follows since $w_{j,\ell}(t)$ appears exactly twice in the sum for every $\{j,\ell\} \neq \{i,i'\}$. Similarly, if $ \{i'\} = P \cap \{i, i'\} $, we can also retrieve $w_{i,i'}(t)$ by computing the bit $D_{\{i\}\cup (P\setminus \{i'\})}^{i'}+\sum_{j\in P\setminus \{i'\}}D_{P\setminus\{j\}}^{j}$. This shows that the user can retrieve the first $2^{N-1}$ bits of $w_{i,i'}$.

For $t\in [2^{N-1}+1:2^{N-1}+2^{N-3}]$, let $ (P_1,P_2) \in \cG $ be the unique set pair such that $ \varphi(P_1,P_2) = t $. By definition, $P_1\subseteq [N]\setminus \{i,i'\}$ and $|P_1|\leq (N-2)/2$. Thus, $P_1\in 2^{N(i')}\setminus \cF$, $(\{i,i'\}\cup P_1\setminus\{j\})\subseteq N(j)$ holds for each $j\in P_1$ and $(\{i,i'\}\cup P_2\setminus\{j\})\subseteq N(j)$ holds for each $j\in P_2$. Then, using the downloaded bits from the servers, the user computes the following bit:  
\begin{align}
     D_{(P_1,P_2)}^{i}&+D_{P_1}^{i'}+D_{P_2}^{i'}+\sum_{j\in P_1}D_{\{i,i'\}\cup P_1\setminus\{j\}}^{j}+ \sum_{j\in P_2}D_{\{i,i'\}\cup P_2\setminus\{j\}}^{j} \nonumber \\
    = & \sum_{\ell \in N(i)} w_{i,\ell}\left({\varphi(P_1,P_2)}\right)+\sum_{\ell \in P_1} w_{i',\ell}\left({\sigma_{i'}(P_1)}\right)+\sum_{\ell \in P_2} w_{i',\ell}\left({\sigma_{i'}(P_2)}\right) \nonumber \\
    & +\sum_{j\in P_1}\sum_{\ell\in \{i,i'\}\cup P_1\setminus\{j\}} w_{j,\ell}\left({\sigma_j(\{i,i'\}\cup P_1\setminus\{j\})}\right) \nonumber \\
    & +\sum_{j\in P_2}\sum_{\ell\in \{i,i'\}\cup P_2\setminus\{j\}} w_{j,\ell}\left({\sigma_j(\{i,i'\}\cup P_2\setminus\{j\})}\right)\label{eq5: Thm scheme rate for K_N}\\
    = & \sum_{\ell \in N(i)} w_{i,\ell}(t)+\sum_{\ell \in P_1} w_{i',\ell}(t) +\sum_{\ell \in P_2} w_{i',\ell}(t) \nonumber \\
    & +\sum_{j\in P_1}\sum_{\ell\in \{i,i'\}\cup P_1\setminus \{j\}} w_{j,\ell}(t) +\sum_{j\in P_2}\sum_{\ell\in \{i,i'\}\cup P_2\setminus \{j\}} w_{j,\ell}(t) \label{eq6: Thm scheme rate for K_N}\\
    = & w_{i,i'}(t)+\sum_{\ell \in [N]\setminus\{i,i'\}} w_{i,\ell}(t)+\sum_{\ell \in [N]\setminus\{i,i'\}} w_{i',\ell}(t) \nonumber \\
    & +\sum_{j\in P_1}\sum_{\ell\in \{i,i'\}\cup P_1\setminus \{j\}} w_{j,\ell}(t) +\sum_{j\in P_2}\sum_{\ell\in \{i,i'\}\cup P_2\setminus \{j\}} w_{j,\ell}(t) r\\
    = & w_{i,i'}(t), \label{eq7: Thm scheme rate for K_N}
\end{align}
where \eqref{eq5: Thm scheme rate for K_N} follows from \eqref{eq8_cons_K_N} and \eqref{eq9_cons_K_N}, \eqref{eq6: Thm scheme rate for K_N} follows from \eqref{eq5_cons_K_N}, \eqref{eq7_cons_K_N}, and the assumption that $ \varphi(P_1, P_2) = t $, and \eqref{eq7: Thm scheme rate for K_N} follows since $ P_1 $ and $ P_2 $ are disjoint sets satisfying $ P_1 \cup P_2 = [N] \setminus \{i, i'\} $, which implies $ w_{j,\ell}(t) $ appears exactly twice in the sum for every $ \{j,\ell\} \neq \{i,i'\} $.

For $t\in [2^{N-1}+2^{N-3}+1:L]$, let $ (P_1,P_2) \in \cG $ be the unique set pair such that $ \varphi(P_1,P_2) = t-2^{N-3} $. Then, using the downloaded bits from the servers, the user computes the following bit:  
\begin{align}
     D_{(P_1,P_2)}^{i'}&+D_{P_1}^{i}+D_{P_2}^{i}+\sum_{j\in [N]\setminus\{i,i'\}}D_{(P_1,P_2)}^{j} \nonumber \\
    = & \sum_{\ell \in N(i')} w_{i',\ell}\left(\varphi(P_1,P_2)+2^{N-3}\right)+\sum_{\ell \in P_1} w_{i,\ell}\left({\sigma_{i}(P_1)}\right) \nonumber \\
    & +\sum_{\ell \in P_2} w_{i,\ell}\left({\sigma_{i}(P_2)}\right) +\sum_{j\in [N]\setminus\{i,i'\}}\sum_{\ell\in N(j)} w_{j,\ell}\left(\varphi(P_1,P_2)+2^{N-3}\right) \label{eq8: Thm scheme rate for K_N}\\
    = & \sum_{\ell \in N(i')} w_{i',\ell}(t)+\sum_{\ell \in P_1} w_{i,\ell}(t) +\sum_{\ell \in P_2} w_{i,\ell}(t) \nonumber \\
    & +\sum_{j\in [N]\setminus\{i,i'\}}\sum_{\ell\in N(j)} w_{j,\ell}(t)  \label{eq9: Thm scheme rate for K_N}\\
    = & w_{i,i'}(t)+\sum_{\ell \in [N]\setminus\{i,i'\}} w_{i',\ell}(t)+\sum_{\ell \in [N]\setminus\{i,i'\}} w_{i,\ell}(t) \nonumber \\
    & +\sum_{j\in [N]\setminus\{i,i'\}}\sum_{\ell\in N(j)} w_{j,\ell}(t) \\
    = & w_{i,i'}(t), \label{eq10: Thm scheme rate for K_N}
\end{align}
where \eqref{eq8: Thm scheme rate for K_N} follows from \eqref{eq8_cons_K_N} and \eqref{eq10_cons_K_N}, \eqref{eq9: Thm scheme rate for K_N} follows from \eqref{eq3_cons_K_N} and the assumption that $ \varphi(P_1, P_2)+2^{N-3} = t $, and \eqref{eq10: Thm scheme rate for K_N} follows since $ w_{j,\ell}(t) $ appears exactly twice in the sum for every $ \{j,\ell\} \neq \{i,i'\} $.

\textbf{Privacy:} Note that, irrespective of the desired file, by \eqref{eq8_cons_K_N}, \eqref{eq9_cons_K_N}, and \eqref{eq10_cons_K_N}, the downloaded $2^{N-1} + 2^{N-3} - 1$ bits from server $S_j$ always have the following feature: one bit for each non-empty set $P$ of $N(j)$, which has the form
\begin{align}
    \sum_{\ell \in P} W_{j,\ell}(*),
\end{align}
and one bit for each set pair $(P_1, P_2) \in \cG$, which has the form
\begin{align}
    \sum_{\ell \in N(j)} W_{j,\ell}(*).
\end{align}
Moreover, by the definition of $w_{j,\ell}$, from the perspective of server $S_j$, the index of the bit from the file $W_{j,\ell}$ in each of the sums above is uniformly distributed over all possible choices. Hence, each server learns no information about the file index. 

\textbf{Rate:} Each server returns $ 2^{N-1}+2^{N-3} - 1 $ bits, so the total number of bits downloaded from all servers is $ N \cdot (2^{N-1}+2^{N-3} - 1) $, while the file size is $ L = 3\cdot 2^{N-2} $. Therefore, the rate is  
\begin{align}
\frac{3\cdot 2^{N-2}}{2^{N-1}+2^{N-3} - 1} \cdot \frac{1}{N}=\frac{6}{5-2^{3-N}}\cdot \frac{1}{N},
\end{align}  
as required.

This completes the proof.
\end{IEEEproof}

\section{Capacity Bounds and Scheme Constructions for PIR over Multigraphs}\label{sec: PIR over multigraphs}

In this section, we investigate the PIR problem over multigraph-based replication systems. As in prior works, we focus on deriving bounds for the PIR capacity and constructing schemes that approach these bounds. We first propose a general PIR scheme for the $r$-multigraph extension of any graph that admits a PIR scheme satisfying a certain symmetry condition. This yields a lower bound on the PIR capacity of multigraphs for which such schemes exist. Then, we establish several other lower and upper bounds on the capacity. Our bounds are shown to be tight for certain classes of multigraphs in specific parameter regimes.

For a simple graph $G = (V, E)$ with $|E| = K'$, we denote by $G^{(r)}$ the \emph{$r$-multigraph extension} of $G$. That is, $G^{(r)}$ is the multigraph defined on the vertex set $V$ by replacing every edge in $G$ with $r$ parallel edges. Thus, $|E(G^{(r)})| = rK' = K$.

Let $\cS = \{S_1, S_2, \ldots, S_N\}$ be a set of $N$ non-colluding servers, and let 
\begin{align}\label{eq_def_file_set_W0}
    \cW_0 = \{W_{1,0}, W_{2,0}, \ldots, W_{K',0}\}
\end{align}
be a collection of $K'$ independent files. This yields a replication system based on the simple graph $G = (\cS, \cW_0)$. Then, the multigraph $G^{(r)} = (\mathcal{S}, \mathcal{W})$, where
\begin{align}\label{eq_def_file_set_W}
    \mathcal{W} = \{W_{i,j} : 1 \leq i \leq K',\, 1 \leq j \leq r\},
\end{align}
characterizes the $2$-replicated system over the server set $\mathcal{S}$ with $rK'$ files indexed by pairs in $[K'] \times [r]$. In this system, each $r$-subset of files $\{W_{\ell,1}, \ldots, W_{\ell,r}\}$ is stored on servers $S_i$ and $S_j$ if and only if $W_{\ell,0}$ is stored on $S_i$ and $S_j$ in the original graph $G$. As in Section \ref{subsec: problem setting}, we denote by $ \cW_{i} $ the set of files stored on server $ S_i $, and by $ \cW_{i,j} $ the set of $r$ files stored on servers $ S_i $ and $ S_j $. Clearly, every $G^{(r)}$ uniquely defines a 2-replication system and the corresponding PIR problem. 

For example, Fig.~\ref{fig2:pir-path-graph} illustrates a system based on the multigraph $ \bbP_3^{(2)} $, i.e., a path graph $ \bbP_3 $ on three vertices where each edge is doubled. The system consists of three servers $ \cS = \{S_1, S_2, S_3\} $ and four files $ \cW = \{W_{1,1}, W_{1,2}, W_{2,1}, W_{2,2}\} $.
 Accordingly, server $S_1$ and $S_3$ store exactly two files, i.e., $ \cW_{1} = \{W_{1,1}, W_{1,2}\} $, $ \cW_{3} = \{W_{2,1}, W_{2,2}\} $, and server $S_2$ stores all the four files, i.e., $ \cW_{2} = \cW $.
\begin{figure}[h]
    \centering
        \begin{tikzpicture}[
            every node/.style={circle, draw, inner sep=0.3pt, minimum size=0.15cm}, 
            every edge/.style={draw, thick}, 
            edge label/.style={draw=none, rectangle, inner sep=2pt, font=\small}
        ]
        \node (A) at (0, 0) {$S_1$};
        \node (B) at (2, 0) {$S_2$};
        \node (C) at (4, 0) {$S_3$};

        \draw[thick, bend right=30] (A) to node[edge label, below] {$W_{1,1}$} (B);
        \draw[thick, bend left=30] (A) to node[edge label, above] {$W_{1,2}$} (B);

        \draw[thick, bend right=30] (B) to node[edge label, below] {$W_{2,1}$} (C);
        \draw[thick, bend left=30] (B) to node[edge label, above] {$W_{2,2}$} (C);
        \end{tikzpicture}
        \caption{The replication system based on $\bbP_3^{(2)}$}
        \label{fig2:pir-path-graph}
\end{figure}

\subsection{A General PIR Scheme for $r$-Multigraphs}\label{subsec: General scheme for r-multigraph}

First, we illustrate the idea of our PIR scheme construction for general $r$-multigraphs using Example~\ref{ex_PIR for P3^{(2)}}. Specifically, based on the scheme for $\bbP_3$ with rate $\frac{2}{3}$ in Example~\ref{ex_PIR for P3}, we construct a scheme for $\bbP_3^{(2)}$ with rate $\frac{2}{3} \cdot \frac{1}{2 - \frac{1}{2}} = \frac{4}{9}$.

\begin{example}\label{ex_PIR for P3^{(2)}}
Consider the PIR problem for the multigraph $ \bbP_3^{(2)} $, as illustrated in Figure \ref{fig2:pir-path-graph}. 

Suppose each file consists of 4 bits, and the user wants to privately retrieve $ W_{\theta} $. The user first privately and independently choose permutations $\sigma_{i,j}:[4]\to [4]$, for $(i,j)\in [K']\times [r]$, uniformly at random to rearrange the bits of each file, where $K'=r=2$. Let the permuted file bits be denoted as:
\begin{align}
& \sigma_{1,1}(W_{1,1}) = (a_1, a_2, a_3, a_4), & & \sigma_{1,2}(W_{1,2}) = (a'_1, a'_2, a'_3, a'_4), \\
& \sigma_{2,1}(W_{2,1}) = (b_1, b_2, b_3, b_4), & & \sigma_{2,2}(W_{2,2}) = (b'_1, b'_2, b'_3, b'_4).
\end{align}
Then, for every possible $\theta$, the user queries and downloads three bits from each server according to Table \ref{tab:multipath_r=2}. 
\begin{table}[h]
\centering
 \begin{tabular}{|c|c|c|c|}   
 \hline
 & $S_1$ & $S_2$ & $S_3$\\
 \hline
 \multirow{3}{*}{$\theta = (1,1)$} & $a_1$ & $a_2+b_2$ & $b_2$\\
 & $a_1'$ & $a_2'+b'_2$ & $b'_2$\\
 & $a_4+a_2'$ & $a_3+a'_1+b_4+b'_4$ & $b_4+b'_4$\\
 \hline
  \multirow{3}{*}{$\theta = (1,2)$} & $a_1$ & $a_2+b_2$ & $b_2$\\
 & $a_1'$ & $a_2'+b'_2$ & $b'_2$\\
 & $a_2+a_4'$ & $a_1+a'_3+b_4+b'_4$ & $b_4+b'_4$\\
 \hline
  \multirow{3}{*}{$\theta = (2,1)$} & $a_1$ & $a_1+b_1$ & $b_2$\\
 & $a_1'$ & $a_1'+b'_1$ & $b'_2$\\
 & $a_2+a_2'$ & $a_2+a'_2+b_4+b'_2$ & $b_3+b'_1$\\
 \hline
  \multirow{3}{*}{$\theta = (2,2)$} & $a_1$ & $a_1+b_1$ & $b_2$\\
 & $a_1'$ & $a_1'+b'_1$ & $b'_2$\\
 & $a_2+a_2'$ & $a_2+a'_2+b_2+b'_4$ & $b_1+b'_3$\\
 \hline
 \end{tabular}
 \vspace{0.2cm}
 \caption{Answer table for $\bbP_3^{(2)}$.}
 \label{tab:multipath_r=2}
\end{table}

For every $\theta = (i,j)$, each server responds to the user's query with three answer bits, indicating that the scheme has rate $\frac{4}{3 \times 3} = \frac{4}{9}$. The first answer bits from all three servers match those in the scheme in Example \ref{ex_PIR for P3}, where the file set is $\{W_{1,1},W_{2,1}\}$ and the desired file is $W_{i,1}$. Similarly, the second answer bits from all three servers match those in the scheme in Example \ref{ex_PIR for P3} with the file set $\{W_{1,2},W_{2,2}\}$ and the desired file $W_{i,2}$. The third answer bits are used to retrieve sums of bits in $W_{i,1}$ and $W_{i,2}$, which are then used to retrieve $W_{i,j}$.

For example, consider the case when $ \theta = (1,1) $. The queries sent to server $S_1$ are:
\begin{align}
    \sigma_{1,1}(1), \sigma_{1,2}(1), \text{ and } \left(\sigma_{1,1}(4), \sigma_{1,2}(2)\right);
\end{align} 
the queries sent to server $S_2$ are:
\begin{align}
    \left(\sigma_{1,1}(2), \sigma_{2,1}(2)\right), \left(\sigma_{1,2}(2), \sigma_{2,2}(2) \right), \text{ and } \left(\sigma_{1,1}(3), \sigma_{1,2}(1), \sigma_{2,1}(4), \sigma_{2,2}(4)\right);
\end{align}
and the queries sent to server $S_3$ are:
\begin{align}
    \sigma_{2,1}(2), \sigma_{2,2}(2), \text{ and } \left(\sigma_{2,1}(4), \sigma_{2,2}(4)\right).
\end{align}
The answer bits from each server are shown in the first row of Table \ref{tab:multipath_r=2}. Using the first and second answer bits from all three servers, the user can retrieve $a_1, a_2$ and $a_1', a_2'$, respectively. From server $S_1$, the third answer bit is a new message bit of $W_{1,1}$ added to a known bit of $W_{1,2}$, i.e., $a_4 + a_2'$. From server $S_3$, the third answer bit is a sum of new message bits of $W_{2,1}$ and $W_{2,2}$, i.e., $b_4 + b_4'$. From $S_2$, the third answer bit is a sum of bits of all four files stored on $S_2$, consisting of a new message bit $a_3$ of $W_{1,1}$, a known bit $a_1'$, and the third answer bit $b_4 + b_4'$, which is also downloaded from $S_3$. Therefore, all four bits of $W_{\theta}=W_{1,1}$ are successfully retrieved. 

To understand why privacy holds, note that the query structure seen by each server is identical for all $ \theta $. Moreover, the private permutations chosen by the user, hide the actual bit indices from each server.
\end{example}

Next, we introduce our general construction of PIR schemes for $r$-multigraphs. Before providing the formal description, we need the following definition.

\begin{definition}[Symmetric Retrieval Property]
A graph-based PIR scheme is said to satisfy the symmetric retrieval property (SRP) if, for every possible $\theta$, the number of bits of the file $W_{\theta}$ retrieved from each server storing $W_{\theta}$ is equal. In other words, for every possible $\theta$, it holds that
\begin{align}
    H(A_{i}|Q_{i}(\theta),\cW_{i}\setminus\{W_{\theta}\})=H(A_{j}|Q_{j}(\theta),\cW_{j}\setminus\{W_{\theta}\})=\frac{H(W_{\theta})}{2},
\end{align}
where $S_i$ and $S_j$ are the servers that store $W_{\theta}$, and $\cW_i$ and $\cW_j$ denote the sets of files stored at $S_i$ and $S_j$, respectively.
\end{definition}

\begin{theorem}\label{Thm: achievable_lbnd}
Let $T$ denote a scheme for the simple graph $G$ with rate $R(G)$ that satisfies SRP. Then, there exists a scheme $T^{(r)}$ for the $r$-multigraph extension $G^{(r)}$ of $G$ with rate $R(G)\cdot(2-2^{1-r})^{-1}$, which provides the following lower bound on the PIR capacity of $G^{(r)}$:
\begin{align}\label{eq:multigraph_achievable_rate}
    \mathscr{C}\left(G^{(r)}\right) &\geq  \frac{R(G)}{2-2^{1-r}}.
\end{align}
\end{theorem}

\begin{remark}\label{rmk1_capacity_LB_multigraph}
    The scheme for paths $\bbP_N$, as described in Construction \ref{cons: scheme for P_N}, satisfies SRP since a single bit of the desired file is retrieved from each server storing it. Then, Theorem \ref{Thm: achievable_lbnd} provides a scheme for $\bbP_N^{(r)}$, leading to  
    \begin{align}
        \mathscr{C}\left(\bbP_N^{(r)}\right) \geq \frac{2}{N} \cdot \frac{1}{2 - 2^{1-r}}.
    \end{align}
    The above bound is tight for even $N$, as demonstrated by a matching upper bound in Theorem \ref{Thm: Capacity upper bound for multigraphs}.
\end{remark}

\begin{remark}\label{rmk2_capacity_LB_multigraph}
    For a vertex-transitive graph, by Lemma 11 in \cite{SGT23}, there exists a PIR scheme that satisfies SRP, and the bound in \eqref{eq:multigraph_achievable_rate} holds. In particular, the optimal scheme for cycle graphs $\bbC_N$ proposed in \cite{BU19} satisfies SRP, yielding the rate $\frac{2}{N+1}$. Thus, Theorem \ref{Thm: achievable_lbnd} implies the following capacity lower bound for multi-cycles:
    \begin{align}
        \mathscr{C}\left(\bbC_N^{(r)}\right) \geq \frac{2}{N+1} \cdot \frac{1}{2 - 2^{1-r}}.
    \end{align}

    One can also verify that the scheme for complete graphs $\bbK_N$ in Construction \ref{cons: scheme for K_N} satisfies SRP, which leads to
    \begin{align}
        \mathscr{C}\left(\bbK_N^{(r)}\right) \geq \frac{6}{N(5 - 2^{3-N})} \cdot \frac{1}{2 - 2^{1-r}}.
    \end{align}
\end{remark}

\begin{remark}\label{rmk3_capacity_LB_multigraph}
    The best-known scheme for star graphs $\bbS_N$ (also denoted by $\bbK_{1,N-1}$), as given in \cite{SGT23}, does not satisfy SRP. However, a trivial scheme, as described next (see also Scheme 1 in \cite{YJ23}), does.

    Let the central server $S_{N}$ store all the files $\{W_1,\ldots,W_{N-1}\}$ where each file consists of $2$ bits, i.e., $W_i = (W_i(1), W_i(2))$. The user privately chooses $N-1$ permutations $\sigma_i: [2] \to [2]$ independently and uniformly at random, and applies $\sigma_i$ to $W_i$ to obtain $w_i \triangleq \sigma_i(W_i) = (w_i(1), w_i(2))$. For any $\theta \in [N-1]$, to retrieve $W_{\theta}$, the user downloads the sum of the first bits of all files, $\sum_{i=1}^{N-1} w_i(1)$, from $S_N$; $w_i(1)$ from $S_i$, where $i \in [N-1] \setminus \{\theta\}$; and $w_{\theta}(2)$ from $S_{\theta}$. Clearly, the scheme satisfies SRP and has the rate $\frac{2}{N}$. Then, by Theorem \ref{Thm: achievable_lbnd}, we have
    \begin{align}
        \mathscr{C}\left(\bbS_N^{(r)}\right) \geq \frac{2}{N} \cdot \frac{1}{2 - 2^{1-r}}.
    \end{align}
\end{remark}
    
\begin{IEEEproof}[Proof of Theorem \ref{Thm: achievable_lbnd}]
Let $T$ be a PIR scheme for the graph $G = (\mathcal{S}, \mathcal{W}_0)$, where $\mathcal{W}_0$ is defined in~\eqref{eq_def_file_set_W0}. Suppose $ T $ admits $L'$ bits per file and $D'$ downloaded bits. Then, $ R(G) = \frac{L'}{D'} $. Next, we construct a PIR scheme $ T^{(r)} $ for $ G^{(r)} $ based on $ T $.

Suppose that each file $W_{s,t}$, $(s,t) \in [K'] \times [r]$, consists of $L = 2^{r-1}L'$ bits in $T^{(r)}$. Before sending out the queries, the user first permutes the $L$ bits of each of the $rK'$ files through a private permutation $\sigma_{s,t}: [L] \rightarrow [L]$, chosen independently and uniformly at random. We denote $\sigma_{s,t}(W_{s,t}) = {w}_{s,t}$. Let $\theta = (i,j)$ be the desired file index, and assume that $W_{i,j}$ is stored on servers $S_{n_1}$ and $S_{n_2}$. Then, as in the motivating example, the scheme $T^{(r)}$ proceeds in $r$ stages. In each stage, we apply $T$ to several graphs on $\cS$, each isomorphic to $G$, where every edge represents a sum of some of the $r$ files stored on its two associated vertices.

In the following, to simplify the description of the retrieval process, we assume without loss of generality that $\theta=(1,1)$.

In the first stage, for every $t\in [r]$, we apply $ T $ to graph $G_{\{t\}}=(\cS, \cW_{\{t\}})$, where 
\begin{align}\label{eq:define_first_stage_graph}
    \cW_{\{t\}}\triangleq\{W_{1,t},W_{2,t},\ldots,W_{K',t}\},
\end{align}
and retrieve the first $ L' $ bits of the file $W_{1,t}$ stored on $ S_{n_1} $ and $ S_{n_2} $. This requires $ rD' $ downloaded bits in total and results in 
\begin{align}
{w}_{1,t}[1:L'],~\forall~t \in [r],
\end{align}
where $ {w}_{1,1}[1:L'] $ are the desired message bits, and $ {w}_{1,t}[1:L'] $, $ t \neq 1 $, are the interference bits used for retrieval in the second stage. By the SRP, we can adjust $ T $ so that, for each $ t \in [r] $, the first half $ {w}_{1,t}[1:L'/2] $ is retrieved from $ S_{n_1} $, and the latter half $ {w}_{1,t}[L'/2+1:L'] $ is retrieved from $ S_{n_2} $.

In the second stage, for every $2$-subset $\{j_1, j_2\} \subseteq [r]$, we apply $T$ to the graph $G_{\{j_1,j_2\}} = (\cS, \cW_{\{j_1,j_2\}})$, where
\begin{equation*}
    \cW_{\{j_1,j_2\}} \triangleq \{W_{i,j_1} + W_{i,j_2} : i \in [K']\},
\end{equation*}
and retrieve $L'$ bits of the file ${W}_{1,j_1} + {W}_{1,j_2}$ stored on $S_{n_1}$ and $S_{n_2}$. Specifically, we perform the following steps:
\begin{itemize}
    \item If $1 \in \{j_1, j_2\}$, assume without loss of generality that $j_1 = 1$. Then, we apply $T$ to retrieve
    \begin{align}
        {w}_{1,1}[(j_2 - 1)L' + 1 : j_2 L'] + {w}_{1,j_2}[1 : L'].
    \end{align}
    \item If $1 \notin \{j_1, j_2\}$, then we apply $T$ to retrieve
    \begin{align}\label{interference_stage2}
        &{w}_{1,j_1}[(j_2 - 1)L' + 1 : j_2 L']  
        +{w}_{1,j_2}[(j_1 - 1)L' + 1 : j_1 L'].
    \end{align}
\end{itemize}
Similarly, by SRP, we can assume that the first $L'/2$ retrieved bits of ${w}_{1,1} + {w}_{1,j_2}$ are from $S_{n_2}$, while the latter $L'/2$ bits are retrieved from $S_{n_1}$. Moreover, for each $\{j_1, j_2\} \subseteq [r] \setminus \{1\}$, the first $L'/2$ retrieved bits of ${w}_{1,j_1} + {w}_{1,j_2}$ are from $S_{n_1}$, and the latter $L'/2$ bits are retrieved from $S_{n_2}$.

Consider the bits of $ {W}_{1,j_1} + {W}_{1,j_2} $ retrieved in the second stage. Since ${w}_{1,j_2}[1:L']$ are known by the first stage, we can retrieve $(r-1)L'$ bits of the form ${w}_{1,1}[(j_2-1)L'+1:j_2L']$, $2\leq j_2\leq r$, for the desired file. Moreover, for each $\{j_1,j_2\}\subseteq [r]\setminus\{1\}$, we also have $L'$ bits from the sum $ {W}_{1,j_1} + {W}_{1,j_2} $. These ${{r-1}\choose 2}L'$ bits are left as interference bits to be used for retrieval in the third stage.

Generally, for $\ell\geq 3$ and every $ (\ell-1) $-set $ B \subseteq [r] \setminus \{1\} $, we uniquely assign it an integer 
\begin{align}
u_{B} \in \left[\sum_{s=0}^{\ell-2}{{r-1}\choose s}+1: \sum_{s=0}^{\ell-2}{{r-1}\choose s} + \binom{r-1}{\ell-1}\right], 
\end{align} 
and denote $U_{B} \triangleq [(u_{B} - 1)L' + 1: u_{B}L']$. Then, in the $\ell$-th stage, for every $ \ell $-subset $ A \subseteq [r] $, $T$ is applied to graph $G_{A}=(\cS,\cW_{A})$, where
\begin{align}
    \cW_{A}\triangleq \left\{\sum_{t\in A}W_{i,t}:~i\in [K']\right\},
\end{align}
to retrieve the following $L'$ bits of $\sum_{t\in A}{W}_{1,t}$: 
\begin{itemize}
    \item If $1\in A$, then we apply $T$ to retrieve
    \begin{align}
    {w}_{1,1}|_{U_{A\setminus\{1\}}}+\sum_{t\in A\setminus\{1\}}{w}_{1,t}|_{U_{A\setminus\{1,t\}}}.
    \end{align}
    \item If $1\notin A$, then we apply $T$ to retrieve \begin{align}\label{interference_stagel}
    \sum_{t\in A}{w}_{1,t}|_{U_{A\setminus\{t\}}}.
    \end{align}
\end{itemize}
By SRP, for $A$ containing $1$, the first $L'/2$ retrieved bits of $\sum_{t\in A} {w}_{1,t}$ are from $S_{n_2}$, while the latter $L'/2$ bits are retrieved from $S_{n_1}$. Moreover, for $A$ not containing $1$, the first $L'/2$ retrieved bits of $\sum_{t\in A} {w}_{1,t}$ are from $S_{n_1}$, while the latter $L'/2$ bits are from $S_{n_2}$.

Using the interference bits of $\sum_{t\in B}{w}_{1,t}|_{U_{B\setminus\{t\}}}$ from the $(\ell-1)$-th stage, we can retrieve ${{r-1}\choose \ell-1}L'$ bits of form ${w}_{1,1}|_{U_{B}}$, for every $B\subseteq [r]\setminus \{1\}$ of size $\ell-1$. For each $A\subseteq [r]\setminus\{1\}$ of size $\ell$, the $L'$ bits from the sum $ \sum_{t\in A}{w}_{1,t}|_{U_{A\setminus\{t\}}} $ are left as interference bits to be used for retrieval in the next stage.

\textbf{Reliability:} The reliability of $T^{(r)}$ follows directly by that of $T$.

\textbf{Privacy:} The privacy of $T^{(r)}$ follows from the privacy of $T$ in each stage and the random permutation on the bits of each file in $\mathcal{W}$. Furthermore, the SRP of $T$ guarantees that the $L'/2$ interference bits of $\sum_{t \in B} {w}_{1,t}|_{U_{B\setminus \{t\}}}$, retrieved from $S_{n_1}$ in the $(\ell-1)$-th stage for some $(\ell-1)$-set $B \subseteq [r] \setminus \{1\}$, are used to help retrieve $L'/2$ message bits of $W_{1,1}$ from ${w}_{1,1}|_{U_B} + \sum_{t \in B} {w}_{1,t}|_{U_{B}}$, which is retrieved from $S_{n_2}$ in the $\ell$-th stage, and vice versa. This ensures that an equal number of bits of every file is downloaded from the perspective of each server.

\textbf{Rate:} For each $\ell\in [r]$, the user downloads $\binom{r}{\ell}D'$ bits in the $\ell$-th stage. The resulting download cost $D$ is therefore,
\begin{align}
    D=\sum_{\ell=1}^r \binom{r}{\ell} D' = (2^r -1)D'.
\end{align}
This gives rise to the achievable rate, $\frac{L}{D} = \frac{2^{r-1}L'}{(2^r-1)D'}$, which is equal to \eqref{eq:multigraph_achievable_rate}.
\end{IEEEproof}

\subsection{Other Capacity Lower Bounds for $r$-Multigraphs}\label{subsec: other lower bounds for multigraphs}

By the PIR reduction results in Section \ref{subsec: PIR reduction}, we can obtain the following simple capacity lower bound for general $r$-multigraphs.

\begin{theorem}\label{Thm: triv_lbnd}
    Let $G$ be a simple graph. Then the PIR capacity of its $r$-multigraph extension $G^{(r)}$ satisfies
    \begin{align}\label{eq:multigraph_triv_lbnd}
        \mathscr{C}(G^{(r)})\geq \frac{\mathscr{C}(G)}{r}.
    \end{align}
\end{theorem}
\begin{IEEEproof}
    The proof follows directly from Theorem \ref{thm-stam} by viewing $G^{(r)}$ as an edge-disjoint union of $r$ isomorphic copies of $G$. 
\end{IEEEproof}

Note that as $r\rightarrow \infty$, the lower bound given by Theorem \ref{Thm: triv_lbnd} approaches $0$, making it trivial. To determine the asymptotic behavior of the PIR capacity $\mathscr{C}(G^{(r)})$ for a given graph $G$, we next present an asymptotic lower bound on $\mathscr{C}(G^{(r)})$, which is a special case of Theorem 1 in \cite{asymp_gxstpir} by Jia and Jafar. 

We define the asymptotic PIR capacity of  $G^{(r)}=(\cS,\cW)$ as:
\begin{align}\label{eq_def_asympto_capacity}
   \mathscr{C}_{\infty} \left(G^{(r)}\right) \triangleq \lim _{r\to \infty} \mathscr{C}\left(G^{(r)}\right).
\end{align}
By \eqref{eq_def_file_set_W}, $\cW$ can be partitioned into disjoint file subsets as: 
$\mathcal{W}=\bigcup_{k=1}^{K'}\mathcal{W}_k$, where $\mathcal{W}_k=\{ W_{k,1}, W_{k,2}, \ldots, W_{k,r}\}$, for each $k\in [K']$. Using the notation of \cite{asymp_gxstpir}, we denote $\mathcal{R}_k$ as the set of servers storing $\mathcal{W}_k$. 
We denote $\rho_k \triangleq |\mathcal{R}_k|$ for each $k\in [K']$ and define
\begin{align}
    \rho_{\min} \triangleq \min_{k\in [K']} \rho_k.
\end{align}
Since we focus on $2$-replicated storage, we have $\rho_k = 2$ for each $k\in [K']$ and $\rho_{\min}=2$. 

During their study of the asymptotic capacity of PIR problems over graphs, Jia and Jafar \cite{asymp_gxstpir} considered a generalized model incorporating additional privacy and security constraints. Specifically, in their setting, the user's privacy is protected against any set of up to $Y$ colluding servers, and the security of the stored data is protected against any set of up to $X$ colluding servers. In this paper, we focus on the setting that does not involve the additional security constraints or server collusion, hence $X=0$ and $Y=1$. Then, the statement of Theorem 1 in \cite{asymp_gxstpir} adopted to our setting yields the following result.
\begin{theorem}\label{asymp_lowerbound_multigraph}\cite[Theorem 1]{asymp_gxstpir}
The asymptotic capacity of $G^{(r)}$ satisfies,
\begin{align}\label{eq:asymp_lbnd_jafar}
    \mathscr{C}_{\infty} \left(G^{(r)}\right) \geq \frac{\rho_{\min}-X-Y}{N} = \frac{1}{N}.
\end{align}
\end{theorem}

As we shall see later in Section \ref{subsec: capacity upper bound for multigraphs}, this lower bound is tight for the $r$-multi-path $\bbP_{N}^{(r)}$ and the $r$-multigraph extension $G^{(r)}$ of any regular graph $G$. Moreover, the above lower bound is achieved by a simple scheme presented in Section III.B of \cite{asymp_gxstpir} (see also the scheme provided by Raviv, Tamo and Yaakobi in \cite[Section III.A]{graphbased_pir}). 

\subsection{Capacity Upper Bounds for PIR over $r$- Multigraphs}\label{subsec: capacity upper bound for multigraphs}

In this subsection, we present two capacity upper bounds for PIR schemes over multigraphs. The first bound applies to general multigraphs and is shown to be optimal for a certain class of graphs. The second bound specifically targets Hamiltonian vertex-transitive graphs and provides a slight improvement over the first. Moreover, both upper bounds are asymptotically tight as $r \to \infty$ by Theorem \ref{asymp_lowerbound_multigraph}.

We begin by recalling some necessary definitions and notations. The \emph{incidence matrix} $I(G)$ of a simple graph $G = (V, E)$ is a $|V(G)| \times |E(G)|$ binary matrix, where the $(i,j)$-th entry is $1$ if and only if the $i$-th vertex is incident with the $j$-th edge. A \emph{matching} $M \subseteq E$ of $G$ is a subset of edges such that no two edges share a common vertex. The \emph{matching number} of $G$, denoted by $\nu(G)$, is the maximum size of a matching in $G$. A graph $ G $ is called \emph{Hamiltonian} if it contains a cycle that visits every vertex in the graph. Moreover, a graph $ G $ is called \emph{Hamiltonian vertex-transitive} if it is both Hamiltonian and vertex-transitive.

Next, we formally state our capacity upper bounds. 

\begin{theorem}\label{Thm: Capacity upper bound for multigraphs}
Let $ G = (\cS, \cW_0) $ be a graph with maximum degree $ \Delta(G) $. Then, the PIR capacity of $G^{(r)}$ satisfies:
\begin{align}
\mathscr{C}\left(G^{(r)}\right) \leq \min \left( \frac{\Delta(G)}{|E(G)|}, \frac{1}{\nu(G)} \right)\cdot\frac{1}{2-2^{1-r}}.
\end{align}
\end{theorem}
\begin{theorem}\label{Thm: multigraph extension of Thm9 in Zachi's paper}
    Let $ G = (\cS,\cW_0) $ be a Hamiltonian vertex-transitive graph. Then, the PIR capacity of $G^{(r)}$ satisfies:
    \begin{align}
        \mathscr{C}(G^{(r)})\leq \frac{1}{N-(N-1)2^{-r}}.
    \end{align}
\end{theorem}

\begin{remark}\label{rmk: Connection between UP bounds of graphs and multigraphs}
    Theorems \ref{Thm: Capacity upper bound for multigraphs} and \ref{Thm: multigraph extension of Thm9 in Zachi's paper} can be viewed as multigraph extensions of Theorems 1 and 9 in \cite{SGT23}, respectively. When $r = 1$, the results of Theorems \ref{Thm: Capacity upper bound for multigraphs} and \ref{Thm: multigraph extension of Thm9 in Zachi's paper} coincide with those of Theorems 1 and 9 in \cite{SGT23}.
\end{remark}

\begin{remark}\label{rmk: Comparsion of UP bounds for multigraphs}
    For finite $r$, the capacity upper bound given by Theorem \ref{Thm: Capacity upper bound for multigraphs} is tight for certain graph classes. For example, when $G = \bbP_N$ and $N$ is even, Theorem \ref{Thm: Capacity upper bound for multigraphs} gives
    \begin{align}
        \mathscr{C}\left(\bbP_N^{(r)}\right) &\leq \min \left( \frac{2}{N-1}, \frac{1}{N/2} \right) \cdot \frac{1}{2 - 2^{1-r}} \\
        &= \frac{2}{N} \cdot \frac{1}{2 - 2^{1-r}}.
    \end{align}
    This upper bound is shown to be tight by Construction \ref{cons: scheme for P_N} and Theorem \ref{Thm: achievable_lbnd}.

    For a regular graph $G$ with $N$ vertices, it always holds that $\frac{\Delta(G)}{|E(G)|} = \frac{2}{N}$. Thus, Theorem \ref{Thm: Capacity upper bound for multigraphs} implies that
    \begin{align}
        \mathscr{C}(G^{(r)}) \leq \frac{1}{N} \cdot \frac{1}{1 - 2^{-r}}
    \end{align}
    holds for any regular graph $G$ with $N$ vertices. Moreover, since every Hamiltonian vertex-transitive graph is a regular graph, Theorem \ref{Thm: multigraph extension of Thm9 in Zachi's paper} provides a slight improvement over this upper bound. Furthermore, as $r \to \infty$, the upper bounds in both Theorems \ref{Thm: Capacity upper bound for multigraphs} and \ref{Thm: multigraph extension of Thm9 in Zachi's paper} reduce to $\frac{1}{N}$, which is tight by Theorem \ref{asymp_lowerbound_multigraph}.
\end{remark}

For the proofs of Theorems \ref{Thm: Capacity upper bound for multigraphs} and \ref{Thm: multigraph extension of Thm9 in Zachi's paper}, we need several preliminary results. To begin with, we prove the following multigraph version of Lemma \ref{Lem3 in Zachi's paper}, which plays a crucial role in the upcoming proofs of the capacity upper bounds for multigraphs.

\begin{lemma}\label{multigraph_version_Lem3}
    Let $S_i$ and $S_j$ be two distinct servers that share a set of $r$ files $\cW_{i,j}\triangleq\{W_1,W_2,\ldots,W_r\}\subseteq \cW$. Then
    \begin{align}
        H(A_i)+H(A_j) & \geq H(A_i|\cW\setminus\cW_{i,j},\cQ)+H(A_j|\cW\setminus\cW_{i,j},\cQ)\\
          & \geq \left(2-\frac{1}{2^{r-1}}\right)L.
    \end{align}
\end{lemma}

For the proof of Lemma~\ref{multigraph_version_Lem3}, we need a modified version of Lemma~6 in~\cite{SJ17}, adapted to the $2$-replicated PIR system in our setting, rather than the original $N$-replicated setting. We state and prove this modified version below. 

\begin{lemma}\label{Lem6 in Sun&Jafar's paper under graph PIR model}
    Let $S_i$ and $S_j$ be two distinct servers that share a set of $r$ files $\cW_{i,j}\triangleq\{W_1,W_2,\ldots,W_r\}\subseteq \cW$. Then for any $s\in \{2,\ldots,r\}$, it holds that
    \begin{align}
         I(\cW_{i,j}&\setminus W_{[s-1]}; A_i,A_j|\cW\setminus\cW_{i,j}\cup W_{[s-1]}, \cQ,\theta=s-1) \nonumber
        \\
        \geq &  \frac{L}{2}+\frac{1}{2}\cdot I(\cW_{i,j}\setminus W_{[s]}; A_i,A_j|\cW\setminus\cW_{i,j}\cup W_{[s]}, \cQ,\theta=s),
    \end{align}
    where $W_{[s]}\triangleq\{W_1,W_2,\ldots,W_s\}$.
\end{lemma}

\begin{IEEEproof}
Denote $\cW^{c}\triangleq \cW\setminus \cW_{i,j}$. For any $s\in \{2,3,\ldots,r\}$, we have
\begin{align}
     2I(\cW_{i,j}&\setminus W_{[s-1]}; A_i,A_j|\cW^{c}\cup W_{[s-1]}, \cQ,\theta=s-1) 
    \nonumber \\
    \geq & I(\cW_{i,j}\setminus W_{[s-1]}; A_i|\cW^{c}\cup W_{[s-1]}, \cQ,\theta=s-1)
    \nonumber \\
    & +I(\cW_{i,j}\setminus W_{[s-1]}; A_j|\cW^{c}\cup W_{[s-1]}, \cQ,\theta=s-1)
     \\
    = & H(A_{i}|\cW^{c}\cup W_{[s-1]}, \cQ,\theta=s-1)-H(A_{i}|\cW,\cQ,\theta=s-1)
    \nonumber \\
    & +H(A_{j}|\cW^{c}\cup W_{[s-1]}, \cQ,\theta=s-1)-H(A_{j}|\cW,\cQ,\theta=s-1)
    \\
    = & H(A_{i}|\cW^{c}\cup W_{[s-1]}, \cQ,\theta=s)-H(A_{i}|\cW,\cQ,\theta=s)
    \nonumber \\
    & +H(A_{j}|\cW^{c}\cup W_{[s-1]}, \cQ,\theta=s)-H(A_{j}|\cW,\cQ,\theta=s)
    \label{eq1_S&J-lem6_graph_model} \\
    = & H(A_{i}|\cW^{c}\cup W_{[s-1]}, \cQ,\theta=s)+H(A_{j}|\cW^{c}\cup W_{[s-1]}, \cQ,\theta=s)
    \label{eq2_S&J-lem6_graph_model} \\
    \geq & H(A_{i}|\cW^{c}\cup W_{[s-1]}, \cQ,\theta=s)+H(A_{j}|\cW^{c}\cup W_{[s-1]}, \cQ,\theta=s,A_{i})
     \\
    = & H(A_{i}|\cW^{c}\cup W_{[s-1]}, \cQ,\theta=s)-H(A_{i}|\cW,\cQ,\theta=s)
    \nonumber \\
    & +H(A_{j}|\cW^{c}\cup W_{[s-1]}, \cQ,\theta=s,A_{i})-H(A_{j}|\cW,\cQ,\theta=s,A_{i})
    \label{eq3_S&J-lem6_graph_model} \\
    = & I(\cW_{i,j}\setminus W_{[s-1]};A_{i},A_{j}|\cW^{c}\cup W_{[s-1]}, \cQ,\theta=s)
    \\
    = & H(\cW_{i,j}\setminus W_{[s-1]}|\cW^{c}\cup W_{[s-1]}, \cQ,\theta=s)
    \nonumber \\
    & -H(\cW_{i,j}\setminus W_{[s-1]}|\cW^{c}\cup W_{[s-1]}, \cQ,\theta=s,A_{i},A_{j})
     \\
    = & H(\cW_{i,j}\setminus W_{[s-1]}|\cW^{c}\cup W_{[s-1]}, \cQ,\theta=s)
    \nonumber \\
    & -H(\cW_{i,j}\setminus W_{[s-1]}|\cW^{c}\cup W_{[s-1]}, \cQ,\theta=s,A_{[N]})
    \label{eq4_S&J-lem6_graph_model} \\
    = & L+H(\cW_{i,j}\setminus W_{[s]}|\cW^{c}\cup W_{[s]}, \cQ,\theta=s)
    \nonumber \\
    & -H(\cW_{i,j}\setminus W_{[s-1]}|\cW^{c}\cup W_{[s-1]}, \cQ,\theta=s,A_{[N]})
    \label{eq5_S&J-lem6_graph_model}  \\
    = & L+H(\cW_{i,j}\setminus W_{[s]}|\cW^{c}\cup W_{[s]}, \cQ,\theta=s)
    \nonumber \\
    & -H(\cW_{i,j}\setminus W_{[s]}|\cW^{c}\cup W_{[s]}, \cQ,\theta=s,A_{[N]})
    \label{eq6_S&J-lem6_graph_model}  \\
    = & L+H(\cW_{i,j}\setminus W_{[s]}|\cW^{c}\cup W_{[s]}, \cQ,\theta=s)
    \nonumber \\
    & ~-H(\cW_{i,j}\setminus W_{[s]}|\cW^{c}\cup W_{[s]}, \cQ,\theta=s,A_i,A_j)
    \label{eq7_S&J-lem6_graph_model} \\
    = & L+I(\cW_{i,j}\setminus W_{[s]}; A_i,A_j|\cW^{c}\cup W_{[s]}, \cQ,\theta=s), 
    \end{align}
where \eqref{eq1_S&J-lem6_graph_model} follows by Proposition \ref{Prop2 in Zachi's paper}, \eqref{eq2_S&J-lem6_graph_model} and \eqref{eq3_S&J-lem6_graph_model} follow by \eqref{eq3_problem_setting}, \eqref{eq4_S&J-lem6_graph_model} and \eqref{eq7_S&J-lem6_graph_model} follow since by \eqref{eq3_problem_setting}, $A_{\ell}$, $\ell\neq i,j$, is a function of $\cW^{c}$ and $\cQ$, \eqref{eq5_S&J-lem6_graph_model} follows by \eqref{eq1_problem_setting} and \eqref{eq2_problem_setting}, and \eqref{eq6_S&J-lem6_graph_model} follows by reliability \eqref{eq4_problem_setting}.
\end{IEEEproof}

Now, with the help of Lemma \ref{Lem6 in Sun&Jafar's paper under graph PIR model}, we can proceed to the proof of Lemma \ref{multigraph_version_Lem3}.

\begin{IEEEproof}[Proof of Lemma \ref{multigraph_version_Lem3}]
    We denote $\cW^{c} \triangleq \cW \setminus \cW_{i,j}$ and $\Delta$ as the following difference of entropies:
    \begin{align}
        \Delta \triangleq & H(\cW_{i,j} \mid \cW^{c}, Q, \theta = 1) -  H(\cW_{i,j} \mid \cW^{c}, \cQ, \theta = 1, A_{[N]})
        \label{eq_Delta_def}.
    \end{align}
    Then, 
    \begin{align}
        \Delta & =H(\cW_{i,j}|\cW^{c}, \cQ,\theta=1)-H(\cW_{i,j}|\cW^{c}, \cQ,\theta=1,A_{i},A_{j})
         \label{eq1_Delta} \\
        & =I(\cW_{i,j}; A_{i},A_{j}|\cW^{c}, \cQ,\theta=1)
          \\
        & =H(A_{i},A_{j}|\cW^{c}, \cQ,\theta=1)-H(A_{i},A_{j}|\cW, \cQ,\theta=1)
          \\
        & =H(A_{i},A_{j}|\cW^{c}, \cQ,\theta=1)
         \label{eq2_Delta} \\
        & \leq H(A_{i}|\cW^{c}, \cQ,\theta=1)+H(A_{j}|\cW^{c}, \cQ,\theta=1) 
          \\
        & =H(A_{i}|\cW^{c}, \cQ)+H(A_{j}|\cW^{c}, \cQ), 
         \label{eq3_Delta}
    \end{align}
    where \eqref{eq1_Delta} and \eqref{eq2_Delta} follow by \eqref{eq3_problem_setting}, and \eqref{eq3_Delta} follows by Proposition \ref{Prop2 in Zachi's paper}.

    On the other hand, we have
    \begin{align}
        \Delta & =rL-H(\cW_{i,j}|\cW^{c}, \cQ,\theta=1,A_{i},A_{j})
         \label{eq4_Delta} \\
        & =rL-\left(H(W_1|\cW^{c}, \cQ,\theta=1,A_{i},A_{j})+H(\cW_{i,j}\setminus\{W_1\}|\cW^{c}\cup W_1, \cQ,\theta=1,A_{i},A_{j})\right)
          \\
        & =rL-\left(H(W_1|\cW^{c}, \cQ,\theta=1,A_{[N]})+H(\cW_{i,j}\setminus\{W_1\}|\cW^{c}\cup W_1, \cQ,\theta=1,A_{i},A_{j})\right)
         \label{eq5_Delta} \\
        & =rL-H(\cW_{i,j}\setminus\{W_1\}|\cW^{c}\cup W_1, \cQ,\theta=1,A_{i},A_{j})
         \label{eq6_Delta} \\ 
        & =rL-\left(H(\cW_{i,j}\setminus\{W_1\}|\cW^{c}\cup W_1, \cQ,\theta=1)-I(\cW_{i,j}\setminus\{W_1\}; A_i,A_j|\cW^{c}\cup W_1, \cQ,\theta=1)\right)
          \\ 
        & =L+I(\cW_{i,j}\setminus\{W_1\}; A_i,A_j|\cW^{c}\cup W_1, \cQ,\theta=1),
         \label{eq7_Delta}
    \end{align}
    where \eqref{eq4_Delta} follows by \eqref{eq1_problem_setting}, \eqref{eq2_problem_setting} and \eqref{eq1_Delta}, \eqref{eq5_Delta} follows by \eqref{eq3_problem_setting}, \eqref{eq6_Delta} follows by reliability \eqref{eq4_problem_setting}, and and \eqref{eq7_Delta} follow by \eqref{eq1_problem_setting} and \eqref{eq2_problem_setting}.

    Next, starting from $s=2$ and applying Lemma \ref{Lem6 in Sun&Jafar's paper under graph PIR model} repeatedly for $s=3$ to $r$, we can obtain that
    \begin{align}
         I(\cW_{i,j}&\setminus\{W_1\}; A_i,A_j|\cW^{c}\cup W_1, \cQ,\theta=1) 
         \nonumber \\
        \geq & \frac{L}{2}+\frac{1}{2}\cdot I(\cW_{i,j}\setminus \cW_{[2]}; A_i,A_j|\cW^{c}\cup \cW_{[2]}, \cQ,\theta=2) 
          \\
        \geq & \frac{L}{2}+\frac{L}{4}+\frac{1}{4}\cdot I(\cW_{i,j}\setminus \cW_{[3]}; A_i,A_j|\cW^{c}\cup \cW_{[3]}, \cQ,\theta=3)
          \\
        \geq & \cdots 
         \nonumber \\
        \geq & \left(\frac{1}{2}+\frac{1}{4}+\cdots+\frac{1}{2^{r-1}}\right)L.
         \label{eq13_Delta}
    \end{align}
    Finally, the result follows by combining \eqref{eq7_Delta} and \eqref{eq13_Delta} together.
\end{IEEEproof}

Now, with the help of Lemma \ref{multigraph_version_Lem3}, we present the proof of Theorem \ref{Thm: Capacity upper bound for multigraphs}, which is similar
to that of \cite[Theorem 1]{SGT23}.

\begin{IEEEproof}[Proof of Theorem \ref{Thm: Capacity upper bound for multigraphs}]
Consider the rate of some PIR scheme $T$ for $ G^{(r)} $:
\begin{align}
    R = \frac{L}{\sum_{i \in [N]} H(A_i)} 
    = \frac{1}{\sum_{i \in [N]} \left(H(A_i)/L\right)} 
    = \frac{1}{\sum_{i \in [N]} \mu_i} 
    = \frac{1}{\mathbf{1}_N \cdot \mu^T},
\end{align}
where $ \mu_i \triangleq \frac{H(A_i)}{L} $ for each $ i \in [N] $, $ \mu \triangleq (\mu_1, \ldots, \mu_N) $, and $ \mathbf{1}_N $ is the all-ones vector of length $ N $. By Lemma \ref{multigraph_version_Lem3}, if servers $ S_i $ and $ S_j $ are adjacent in $G^{(r)}$, then $\mu_i + \mu_j\geq 2-\frac{1}{2^{r-1}}$. Hence, we can get an upper bound on $R$ via the reciprocal of the optimal value of the following linear program:
\begin{align}
     \min \quad & \mathbf{1}_N \cdot \mu^T, \nonumber \\
    \text{s.t.} \quad & I(G)^T \cdot \mu^T \geq \left( 2-\frac{1}{2^{r-1}} \right) \cdot \mathbf{1}_{K'}; \nonumber \\
    & \mu_i \geq 0,~\forall~i\in [N],
\end{align}
where $ I(G) $ is the incidence matrix of graph $ G $ and $K'=|\cW_0|$. Its dual problem is
\begin{align}
    \max  \quad &  \left( 2-\frac{1}{2^{r-1}} \right) \cdot\mathbf{1}_{K'} \cdot \eta^T, \nonumber \\
    \text{s.t.} \quad & I(G) \cdot \eta^T \leq \mathbf{1}_N; \nonumber \\
    & \eta_i  \geq 0,~\forall~i\in [K'],
\end{align}
where $ \eta $ is a vector of length $ K' $. By the primal-dual theory, any feasible solution to the dual problem provides a lower bound for the primal problem. Below, we provide two feasible solutions to the dual problem:

\begin{itemize}
    \item [(S1)] One can easily verify that $ \bbv_1 = \frac{1}{\Delta(G)} \cdot \mathbf{1}_{K'} $ is a feasible solution of the dual problem, where $ \Delta(G) $ is the maximum degree of $ G $. Therefore, by
    \begin{align}
    \mathbf{1}_{K'} \cdot \bbv_1^T = \frac{K'}{\Delta(G)},
    \end{align}
    the rate $R$ is at most $ \Delta(G)/K' \cdot \left(2-2^{1-r} \right)^{-1}$.
    \item [(S2)] Let $ M \subseteq \cW_{0} $ be a maximum matching of $ G $, i.e., $ |M| = \nu(G) $. Let $ \bbv_2 \in \{0, 1\}^{K'} $ be the indicator vector of $ M $, i.e., $ \bbv_2(i) = 1 $ if and only if the $i$-th file $ W_{i,0} $ is contained in $ M $. Again, it is easy to verify that $ \bbv_2 $ is a feasible solution of the dual problem, and therefore, the rate $R$ is at most $ 1/\nu(G)\cdot \left(2-2^{1-r} \right)^{-1} $.
\end{itemize}

This completes the proof.    
\end{IEEEproof}

With the help of Lemma \ref{multigraph_version_Lem3}, the proof of Theorem \ref{Thm: multigraph extension of Thm9 in Zachi's paper} also follows similarly to that of Theorem 9 in \cite{SGT23}. Thus, we do not repeat it in the main text and instead defer it to the Appendix \ref{appendix: pf of Thm IV.8} for readers' convenience.

\section{Conclusion and Future Directions}\label{sec: conclusion}

In this paper, we studied the PIR problem over graph- and multigraph-based replication systems with non-colluding servers. We established several upper bounds on the PIR capacity for specific classes of graphs and general $r$-multigraphs, which were shown to be tight in certain cases. Additionally, we provided constructions of PIR schemes for some graph classes and general $r$-multigraphs satisfying certain conditions, improving existing PIR schemes and leading to enhanced capacity lower bounds.

Despite these results, the diverse graph structures that can model replication systems in practical scenarios leave many intriguing problems widely open in this topic. Below, we highlight several open questions that we believe will contribute to a deeper understanding of PIR problems in (hyper-)graph- and multigraph-based replication systems:
\begin{enumerate}
    \item In Section \ref{subsec: PIR over P_N}, we showed that the PIR capacity of the path graph $\mathbf{P}_N$ is $\frac{2}{N}$, and in Section \ref{subsec: PIR over K_M,N}, we proved that the PIR capacity of the complete bipartite graph $\mathbf{K}_{M,N}$ satisfies \eqref{eq_PIR capacity of K_M,N}. These results, together with the capacity bounds for cycles, stars, and complete graphs from previous works \cite{graphbased_pir,BU19,SGT23,YJ23}, lead us to question whether graphs containing long paths necessarily exhibit smaller PIR capacities, while graphs without long paths tend to have larger PIR capacities. Resolving this question, whether by proving or disproving it, would provide insights into the PIR capacities of general graphs.
    \item In Section \ref{subsec: General scheme for r-multigraph}, we showed in Theorem \ref{Thm: achievable_lbnd} that for any graph, a PIR scheme can always be constructed for its $r$-multigraph extension, provided that the original graph admits a PIR scheme satisfying SRP. This raises the question of whether this condition can be removed or relaxed to achieve a more general characterization of the PIR capacities of a graph and its $r$-multigraph extension. Alternatively, can it be shown that every graph admits a capacity-achieving PIR scheme that satisfies SRP?
    \item In Section \ref{subsec: capacity upper bound for multigraphs}, we extended two known PIR capacity upper bounds for graphs, stated as Theorem 1 and Theorem 9 in \cite{SGT23}, to the $r$-multigraph case. However, there is another capacity upper bound, Theorem 6 in \cite{SGT23}, which we were unable to generalize. Note that Theorem 6 in \cite{SGT23} was used to establish the capacity upper bounds for $\bbP_N$ and $\bbK_{M,N}$ in Section \ref{subsec: PIR over P_N} and Section \ref{subsec: PIR over K_M,N}, respectively. We believe an $r$-multigraph extension of Theorem 6 in \cite{SGT23} could also be helpful for getting better capacity upper bounds for certain $r$-multigraphs.
    \item In the graph- and multigraph-based replication systems considered in this paper, each file is assumed to be stored on exactly two servers, which is a restrictive condition. Therefore, another natural question is whether the results in this paper, as well as those from previous works, can be extended to hypergraph-based replication systems.
\end{enumerate}

\appendices

\section{Proofs of Theorem \ref{thm-stam}}\label{appendix: PIR reduction}

\begin{IEEEproof}[Proof of Theorem~\ref{thm-stam}]
     Let $T_1$ and $T_2$ be PIR schemes for $H_1$ and $H_2$, respectively, both with the same file length $L$ and rates $R_1$ and $R_2$. Next, by running $T_1$ and $T_2$ independently, we construct a PIR scheme for $G$ with rate $\left(R_1^{-1}+R_2^{-1}\right)^{-1}$ and file length $L$. Then, \eqref{stam} follows directly from the definition of the PIR capacity.

    We assume without loss of generality that $V(G) = V_1 = V_2 = [N]$. Then, we define the PIR scheme for $G$ as follows:
    \begin{itemize}
        \item [(a)] The user chooses a pair of file indices $(\theta_{1},\theta_2)$ from $[|E_{1}|]\times [|E_{2}|]$ uniformly at random and conceals the actual desired file index as one of $\theta_1$ or $\theta_2$.

        \item [(b)] The user sends a query $ Q_i(\theta_1,\theta_2) = \big(Q_{i}^{T_{1}}(\theta_{1}), Q_{i}^{T_{2}}(\theta_{2})\big) $ to each server $ S_i,~i \in [N] $, where $ Q_{i}^{T_{j}}(\theta_{j}) $, $ j \in [2] $, represents the query sent to server $ S_i $ in the scheme $ T_j $ with desired file index $ \theta_j $.

        \item [(c)] Each server $ S_i,~i \in [N]$ returns the answer $A_i$, defined as: 
        \begin{align}
            A_i(\theta_1,\theta_2) = \left(A_{i}\left(Q_{i}^{T_{1}}(\theta_{1})\right), A_{i}\left(Q_{i}^{T_{2}}(\theta_{2})\right)\right),
        \end{align}
        where $A_{i}\left(Q_{i}^{T_{j}}(\theta_{j})\right)$, $j \in [2]$, is the answer from server $S_i$ in the scheme $T_j$ upon receiving the query $Q_{i}^{T_{j}}(\theta_{j})$.
    \end{itemize}
    
    We now prove that this scheme has the claimed rate and satisfies the privacy and reliability requirements.

    \textbf{Rate:} Since $ T_1 $ and $ T_2 $ have the same file length $ L $ and rates $ R_1 $ and $ R_2 $, respectively, we have
    \begin{align}
        \sum_{i \in [N]} H\left(A_i\left(Q_{i}^{T_{j}}(\theta_{j})\right)\right) = \frac{L}{R_j},
    \end{align}
    for each $ j \in [2] $. Moreover, since $ T_1 $ and $ T_2 $ are independent schemes and $ \theta_1 $ and $ \theta_2 $ are chosen independently at random, it follows that
    \begin{align}
        \sum_{i \in [N]} H(A_i) = \sum_{i \in [N]} \sum_{j \in [2]} H\left(A_i\left(Q_{i}^{T_{j}}(\theta_{j})\right)\right) = \frac{L}{R_1} + \frac{L}{R_2}.
    \end{align}
    Thus, the proposed scheme has rate $ \left(R_1^{-1} + R_2^{-1}\right)^{-1} $.

    \textbf{Privacy:} For each $i \in [N]$, let $\cW_i$ denote the set of files stored on server $S_i$. For each $j \in [2]$, let $\cW_{i,j}$ be the set of files in $H_j$ that are stored on $S_i$. Clearly, $\cW_{i,1}$ and $\cW_{i,2}$ are disjoint, and we have $\cW_i = \cW_{i,1} \cup \cW_{i,2}$. 

    Since $T_1$ and $T_2$ satisfy the privacy requirement, by \eqref{eq5_problem_setting}, for each $i \in [N]$ and $j \in [2]$, it holds that
    \begin{align}
        \left(Q_i^{T_j}(\theta_j),A_{i}\left(Q_{i}^{T_j}(\theta_j)\right),\cW_{i,j}\right) &\sim \left(Q_i^{T_j}(1),A_{i}\left(Q_{i}^{T_j}(1)\right),\cW_{i,j}\right).
    \end{align}
    This leads to 
    \begin{align}
        \left(Q_i(\theta_1,\theta_2),A_{i}(\theta_1,\theta_2),\cW_{i}\right) \sim \left(Q_i(1,1),A_{i}(1,1),\cW_{i}\right),
    \end{align}
    which implies that the above scheme for $G$ satisfies the privacy requirement.

    \textbf{Reliability:} Based on the answers returned from the servers, for each $j \in [2]$, the user can obtain the answer set $\left\{A_i\left(Q_{i}^{T_{j}}(\theta_{j})\right) : i \in [N]\right\}$. By the reliability of $T_1$ and $T_2$, the user can retrieve both the $\theta_1$-th file in $H_1$ and the $\theta_2$-th file in $H_2$, this ensures reliability of the proposed scheme.

    Finally, we complete the proof for this case by showing why one can assume without loss of generality that schemes $T_1$ and $T_2$ have the same file length. Suppose that $T_1$ and $T_2$ have different file lengths $L_1$ and $L_2$, respectively. Clearly, for any integer $k\geq 1$, one can obtain a scheme $T_1'$ for $H_1$ with the same rate and file length $kL_1$ by running the scheme $T_1$ independently $k$ times. Through this approach, we can obtain schemes $T_1'$ and $T_2'$ for $H_1$ and $H_2$ with the same file length $L' = \text{lcm}(L_1, L_2)$, respectively, while maintaining their rates. 
    
    Next, we show that equality holds in \eqref{stam} if $G$ is an vertex-disjoint union of $H_1$ and $H_2$.

    We assume without loss of generality that $V(G) = [N]$, $V_1=[N_1]$ and $V_2=[N]\setminus [N_1]$ for some integer $1\leq N_1<N$. Suppose, towards a contradiction, that there exists a PIR scheme $T$ for $G$ with file length $L$ and rate $R > \left(\mathscr{C}(H_1)^{-1}+\mathscr{C}(H_2)^{-1}\right)^{-1}$. Hence, for a desired file index $\theta$ chosen uniformly at random from $[|E(G)|]$, the answer $A_i^{T}$ from server $S_i$ satisfies
    \begin{align}\label{eq1: Thm vertex-disjoint reduction}
        \sum_{i \in [N]} H(A_i^{T}(\theta)) &= \frac{L}{R} < L\left(\mathscr{C}(H_1)^{-1}+\mathscr{C}(H_2)^{-1}\right).
    \end{align}
    Meanwhile, by the privacy requirement \eqref{eq5_problem_setting}, we know that
    \begin{align}\label{eq2: Thm vertex-disjoint reduction}
        A_{i}^{T}(1)\sim A_{i}^{T}(\theta)
    \end{align}
    holds for each $i\in [N]$.
    
    By running the scheme $T$ only over the servers $\{S_1, \ldots, S_{N_1}\}$ and the file set $\bigcup_{i \in [N_1]} \cW_{i}$, we obtain a PIR scheme $T_1$ for $H_1$. Specifically, in $T_1$, for a desired file index $\theta_1$ chosen uniformly at random from $[|E(H_1)|]$, each server $S_i$, where $i \in [N_1]$, receives a query $Q_{i}^{T}(\theta_1)$ and returns the answer $A_i^{T}(\theta_1)$. Clearly, $T_1$ has file length $L$ and rate $R_1 \leq \mathscr{C}(H_1)$. Hence, the answer $A_i^{T}(\theta_1)$ from the server $S_i$ satisfies  
    \begin{align}\label{eq3: Thm vertex-disjoint reduction}  
        \sum_{i \in [N_1]} H(A_i^{T}(\theta_1)) &= \frac{L}{R_1} \geq \frac{L}{\mathscr{C}(H_1)}.  
    \end{align}

    Similarly, by running the scheme $T$ only over the servers $\{S_{N_1+1}, \ldots, S_{N}\}$ and the file set $\bigcup_{i \in [N]\setminus[N_1]} \cW_{i}$, we obtain a PIR scheme $T_2$ for $H_2$ with file length $L$ and rate $R_2 \leq \mathscr{C}(H_2)$, which leads to
    \begin{align}\label{eq4: Thm vertex-disjoint reduction}
        \sum_{i \in [N]\setminus [N_1]} H(A_i^{T}(\theta_2)) &=  \frac{L}{R_2}\geq \frac{L}{\mathscr{C}(H_2)},
    \end{align}
    for some $\theta_2$ chosen uniformly at random from $[|E(G)|]\setminus [|E(H_1)|]$.

    Then, by \eqref{eq2: Thm vertex-disjoint reduction}, \eqref{eq3: Thm vertex-disjoint reduction} and \eqref{eq4: Thm vertex-disjoint reduction} together lead to 
    \begin{align}
        L\left(\mathscr{C}(H_1)^{-1}+\mathscr{C}(H_2)^{-1}\right) & \leq \sum_{i \in [N_1]} H(A_i^{T}(\theta_1))+\sum_{i \in [N]\setminus [N_1]} H(A_i^{T}(\theta_2)) \\
        & = \sum_{i \in [N]} H(A_i^{T}(1)),
    \end{align}
    which contradicts \eqref{eq1: Thm vertex-disjoint reduction}. This completes the proof.
\end{IEEEproof}

\section{Proof of Lemma \ref{Variant of Lem11 in Zachi's paper}}\label{pf of variant of Lem11 in Zachi's paper}

Let $T$ be a PIR scheme for a vertex-part-transitive bipartite graph $G$ with vertex bipartition $\{U, V\}$. Recall that $\Gamma$ is the subgroup of $\mathrm{Aut}(G)$ that acts transitively on both $U$ and $V$. Then, given an automorphism $f\in \Gamma$, one can construct the scheme $T_{f}$ to retrieve the file with index $\theta$ as follows. 

Suppose $W_{\theta}$ is stored on servers $S_u$ and $S_v$. Denote $S_{f(i)}$ as the image of $S_{i}$ after applying $f$ to $G$ and let $f(\theta)$ be the index of the unique file stored on servers $S_{f(u)}$ and $S_{f(v)}$. Then, to server $S_i$, the user sends the query
\begin{align}\label{eq1_pf_variant_lem11}
    Q_{i}^{T_f} \triangleq Q_{f(i)}^{T}(f(\theta)),
\end{align}
and the server replies with the answer:]
\begin{align}\label{eq2_pf_variant_lem11}
    A_{i}^{T_f} \triangleq A_{f(i)}^{T}.
\end{align}

According to (\ref{eq1_pf_variant_lem11}) and (\ref{eq2_pf_variant_lem11}), the scheme $T_f$ can be viewed as applying the scheme $T$ on the graph $f(G)$ with the desired file $W_{f(\theta)}$. Note that the graph $f(G)$ is simply a relabeling of every vertex (server) $S_i$ of $G$ by $S_{f(i)}$. By the reliability of $T$, the user can recover the file $W_{f(\theta)}$, which is the unique file stored on $S_{f(u)}$ and $S_{f(v)}$. Since $f$ is just a relabeling of vertices, it follows that this is also the unique file stored on servers $S_u$ and $S_v$ in the graph $G$, i.e., the file indexed by $\theta$, as required. The privacy requirement of $T_f$ follows since we are simply running the original scheme $T$, which satisfies privacy. Clearly, the rate of $T_f$ is the same as that of $T$.

Next, we present the proof of Lemma \ref{Variant of Lem11 in Zachi's paper}

\begin{IEEEproof}[Proof of Lemma \ref{Variant of Lem11 in Zachi's paper}]
To establish a PIR scheme with the desired property, we define a scheme $T'$ by uniformly selecting one of the schemes $T_f$, for $f \in \Gamma$, and using it to retrieve the desired file as follows:
\begin{itemize}
    \item [(a)] The user selects a file $\theta \in [K]$ and an automorphism $f \in \Gamma$ independently and uniformly at random.
    \item [(b)] The user generates the queries $\cQ^{T'} \triangleq \cQ^{T_f}=\{Q_i^{T_f}, i\in [N]\}$ and sends them to the servers along with the automorphism $f$.
    \item [(c)] Each server $S_i$ responds with the answer $A_{i}^{T_f}$.
\end{itemize}

Clearly, $T'$ is a scheme with the same rate as $T$. Let $S_i$ and $S_{i'}$ be two vertices in part $U$. By the transitivity of $\Gamma$, there exists an automorphism $g \in \Gamma$ of $G$ such that $S_{g(i')} = S_{i}$. Then:
\begin{align}
    H(A_{i}^{T'}|\cQ^{T'}) &= \frac{1}{|\Gamma|}\sum_{f \in \Gamma}H(A_{i}^{T_f}|\cQ^{T_f}) \label{eq3_pf_variant_lem11}\\
    &= \frac{1}{|\Gamma|}\sum_{f \in \Gamma}H(A_{f(i)}^{T}|\cQ^{T})  \\
    &= \frac{1}{|\Gamma|}\sum_{f \in \Gamma}H(A_{f(g(i'))}^{T}|\cQ^{T})  \\
    &= \frac{1}{|\Gamma|}\sum_{f \in \Gamma}H(A_{f(i')}^{T}|\cQ^{T}) \label{eq4_pf_variant_lem11}\\
    &= \frac{1}{|\Gamma|}\sum_{f \in \Gamma}H(A_{i'}^{T_f}|\cQ^{T_f})  \\
    &= H(A_{i'}^{T'}|\cQ^{T'}), 
\end{align}
where equation (\ref{eq3_pf_variant_lem11}) follows from the fact that $H(X) = \mathbb{E}[-\log(p(X))]$ and that $f \in \Gamma$ is chosen uniformly, and equation (\ref{eq4_pf_variant_lem11}) follows because if $f$ ranges over all automorphisms in $\Gamma$, then so does $f \circ g$. This proves that $H(A_i|\cQ) = H(A_{i'}|\cQ)$ for any $S_i,S_{i'} \in U$.

Similarly, one can show that $H(A_j|\cQ) = H(A_{j'}|\cQ)$ for any $S_j,S_{j'} \in V$. This concludes the proof.
\end{IEEEproof}

\section{Proof of Theorem \ref{Thm: multigraph extension of Thm9 in Zachi's paper}}\label{appendix: pf of Thm IV.8}

\begin{IEEEproof}[Proof of Theorem \ref{Thm: multigraph extension of Thm9 in Zachi's paper}]
    Let $\cW$ denote the set of all files over $G^{(r)}$. Let $ (S_1, \cW_1, S_2, \cW_2, \dots, S_N, \cW_N, S_1) $ be a Hamiltonian $r$-multicycle in $ G^{(r)} $, where $\cW_{i}=\{W_{i,1},W_{i,2},\ldots,W_{i,r}\}$ is the set of $r$ files stored on servers $S_i$ and $S_{i+1}$, modulo $N$. Assume that each file has length $L$ and $\theta=(N,1)$ is the desired file index. Then,  we have
    \begin{align}
    L= H(W_{N,1}) \leq & H(\cW_N \mid  \cQ, \theta = (N,1) ) - H(\cW_N \mid A_{[N]}, \cQ, \theta = (N,1)) \label{eq1_Thm extension of Zachi's Thm9} \\
    = & I(\cW_N; A_{[N]} \mid \cQ, \theta = (N,1))  \\
    = & H(A_{[N]} \mid \cQ, \theta = (N,1)) - H(A_{[N]} \mid \cW_N, \cQ, \theta = (N,1))  \\
    \leq & \sum_{i\in [N]}H(A_i \mid \cQ, \theta = (N,1)) - H(A_{[N]} \mid \cW_N, \cQ, \theta = (N,1))  \\
    \leq & \sum_{i\in [N]}H(A_i) - H(A_{[N]} \mid \cW_N, \cQ, \theta = (N,1)) \label{eq2_Thm extension of Zachi's Thm9},
    \end{align}
    where \eqref{eq1_Thm extension of Zachi's Thm9} follows since $H(\cW_N \mid \cQ, \theta = (N,1))= rL$ and $H(\cW_N \mid A_{[N]}, \cQ, \theta = (N,1))\leq (r-1)L$ by reliability \eqref{eq4_problem_setting}.
    
    Since the proof of Lemma 11 in \cite{SGT23} can be directly extended to $r$-multigraphs, it follows from Lemma 11 in \cite{SGT23} and Lemma \ref{multigraph_version_Lem3} that
    \begin{align}
        H(A_{i} \mid \cW \setminus \cW_{i}, \cQ, \theta = (i,j)) \geq (1 - 2^{-r})L \label{eq3_Thm extension of Zachi's Thm9}
    \end{align}
    holds for each $i \in [N], j\in [r]$. Then, by rearranging inequality \eqref{eq2_Thm extension of Zachi's Thm9}, we have
    \begin{align}
    \sum_{i\in [N]}H(A_i) \geq & L+ H(A_{[N]} \mid \cW_N, \cQ, \theta = (N,1))  \\
    \geq & L+ H(A_{[N-1]} \mid \cW_N, \cQ, \theta = (N,1))  \\
    = & L+ \sum_{i=1}^{N-1} H(A_{i} \mid \cW_N, A_{[i-1]}, \cQ, \theta = (N,1)) \label{eq4_Thm extension of Zachi's Thm9} \\
    \geq &  L+ \sum_{i=1}^{N-1} H(A_{i} \mid \cW_N, A_{[i-1]}, \cW\setminus \cW_{i} , \cQ, \theta = (N,1))
    \\
    \geq & L+ \sum_{i=1}^{N-1} H(A_{i} \mid \cW\setminus \cW_{i}, \cQ, \theta = (N,1)) \label{eq5_Thm extension of Zachi's Thm9} \\
    \geq & L+(N-1)(1-2^{-r})L \label{eq6_Thm extension of Zachi's Thm9} \\
    = & \left(N-\frac{N-1}{2^r}\right)L. \label{eq7_Thm extension of Zachi's Thm9}
    \end{align}
    where \eqref{eq4_Thm extension of Zachi's Thm9} follows by the chain rule of entropy, \eqref{eq5_Thm extension of Zachi's Thm9} follows since the file set $\cW_i$ is stored only on servers $S_i$ and $S_{i+1}$ and by \eqref{eq3_problem_setting}, the answers $A_{[i-1]}$ are therefore determined by $\cQ$ and $\cW \setminus \cW_i$, and \eqref{eq6_Thm extension of Zachi's Thm9} follows by \eqref{eq3_Thm extension of Zachi's Thm9}.

    Finally, the results follows directly by $R = \frac{L}{\sum_{i\in [N]}H(A_i)}$ and \eqref{eq7_Thm extension of Zachi's Thm9}.
\end{IEEEproof}

\bibliographystyle{IEEEtran}
\bibliography{biblio_new}
\end{document}